%% file: 00-main.tex
\PassOptionsToPackage{usenames,dvipsnames}{xcolor}
\documentclass[acmtog]{acmart}

\input{01-acm}
\input{02-package}
\input{03-macros}
\input{04-symbols}

\begin{document}

\title{Topology-First B-Rep Meshing}

\author{YunFan Zhou}
\affiliation{%
  \institution{New York University}
  \country{USA}
}
\author{Daniel Zint}
\affiliation{%
  \institution{New York University}
  \country{USA},
  \institution{TU Berlin}
  \country{Germany}
}
\author{Nafiseh Izadyar}
\affiliation{%
  \institution{University of Victoria}
  \country{Canada}
}
\author{Michael Tao}
\affiliation{%
  \institution{New York University}
  \country{USA}
}
\author{Daniele Panozzo}
\affiliation{%
  \institution{New York University}
  \country{USA}
}
\author{Teseo Schneider}
\affiliation{%
  \institution{University of Victoria}
  \country{Canada}
}

\date{}

\input{figs/teaser}

\begin{abstract}
	\input{10-abstract}

\end{abstract}

\maketitle



\input{11-intro}
\input{12-background-daniele}

\input{20-related}
\input{30-method}
\input{40-implementation}

\input{50-results}
\input{60-conclusion}


\bibliographystyle{ACM-Reference-Format}
\bibliography{99-biblio,98-chatgpt}

\appendix
\input{90-appendix}

\end{document}

%% file: 01-acm.tex
\setcopyright{none}
\renewcommand\footnotetextcopyrightpermission[1]{}

%% file: 02-package.tex
\usepackage[l2tabu,orthodox]{nag}

\usepackage[usenames,dvipsnames]{xcolor} 
\usepackage{amsthm} 

\usepackage[normalem]{ulem}

\usepackage{amsthm}
\usepackage{amsmath}
\usepackage{amsfonts}
\usepackage{mathtools}
\usepackage{stmaryrd}
\usepackage{textcomp} 
\usepackage{siunitx}
\usepackage{diffcoeff}
\usepackage{scalerel}

\usepackage[T1]{fontenc}
\usepackage[utf8]{inputenc}

\usepackage{xparse}
\usepackage{geometry}
\usepackage{ifthen}
\usepackage{float} 
\usepackage{hyperref}
\usepackage{cleveref}
\usepackage{csquotes}
\usepackage{calc}
\usepackage[usenames,dvipsnames]{xcolor}
\usepackage{natbib}
\usepackage{subcaption}
\usepackage{caption}

\usepackage{algorithm}
\usepackage{algorithmicx}
\usepackage{algpseudocode}
\usepackage{listingsutf8}

\usepackage{array}
\usepackage{booktabs}
\usepackage{multirow}

\usepackage{float}
\usepackage{wrapfig} 
\usepackage{graphicx}
\usepackage[percent]{overpic}
\usepackage{caption}
\usepackage{subcaption}
\usepackage{tikz}
\usetikzlibrary{decorations.pathmorphing,patterns,shadows,calc}



%% file: 03-macros.tex
\ExplSyntaxOn
\DeclareExpandableDocumentCommand{\IfNoValueOrEmptyTF}{mmm}
 {
  \IfNoValueTF{#1}{#2}
   {
    \tl_if_empty:nTF {#1} {#2} {#3}
   }
 }
\ExplSyntaxOff

\definecolor{DanieleColor}{RGB}{231,41,138}
\definecolor{DnznColor}{RGB}{217,95,2}
\definecolor{JeremieColor}{RGB}{117,112,179}
\definecolor{MarcosColor}{RGB}{231,41,138}
\definecolor{PiereColor}{RGB}{102,166,30}
\definecolor{TamyColor}{RGB}{230,171,2}
\definecolor{YunFanColor}{RGB}{166,118,29}
\definecolor{MTaoColor}{RGB}{117,112,240}
\definecolor{teseoCol}{rgb}{.15, .68, .38}

\providecommand{\finalversion}{0} 
\ifthenelse{\equal{\finalversion}{1}}
{
	\renewcommand{\jeremie}[1]{}
	\renewcommand{\padawan}[1]{}
	\renewcommand{\toref}[1]{}
	\renewcommand{\tocite}[1]{}
	\renewcommand{\todo}[1]{}
	\renewcommand{\warning}[1]{}
	\renewcommand{\note}[1]{}
}
{}

\definecolor{forestgreen}{rgb}{0.13,0.54,0.13}
\definecolor{darkblue}{rgb}{0,0,.5}
\hypersetup{
	unicode=true,
	colorlinks=true,
	citecolor=forestgreen, 
	linkcolor=darkblue, 
	urlcolor=darkblue, 
}

\newcommand{\filename}[1]{\url{#1}}
\newcommand{\foldername}[1]{\url{#1}}

\algrenewcommand{\algorithmiccomment}[1]{{\footnotesize\color{forestgreen}\(\triangleright\) #1}}

\crefname{algocf}{alg.}{algs.}
\Crefname{algocf}{Algorithm}{Algorithms}

\crefname{appsec}{Appendix}{Appendices}



%% file: 04-symbols.tex
\let\originalleft\left
\let\originalright\right
\renewcommand{\left}{\mathopen{}\mathclose\bgroup\originalleft}
\renewcommand{\right}{\aftergroup\egroup\originalright}

\DeclarePairedDelimiterX{\inner}[2]{\langle}{\rangle}{#1, #2}



%% file: figs/teaser.tex
\begin{teaserfigure}
\footnotesize
 \vspace{-1em}
 \includegraphics[width=\textwidth]{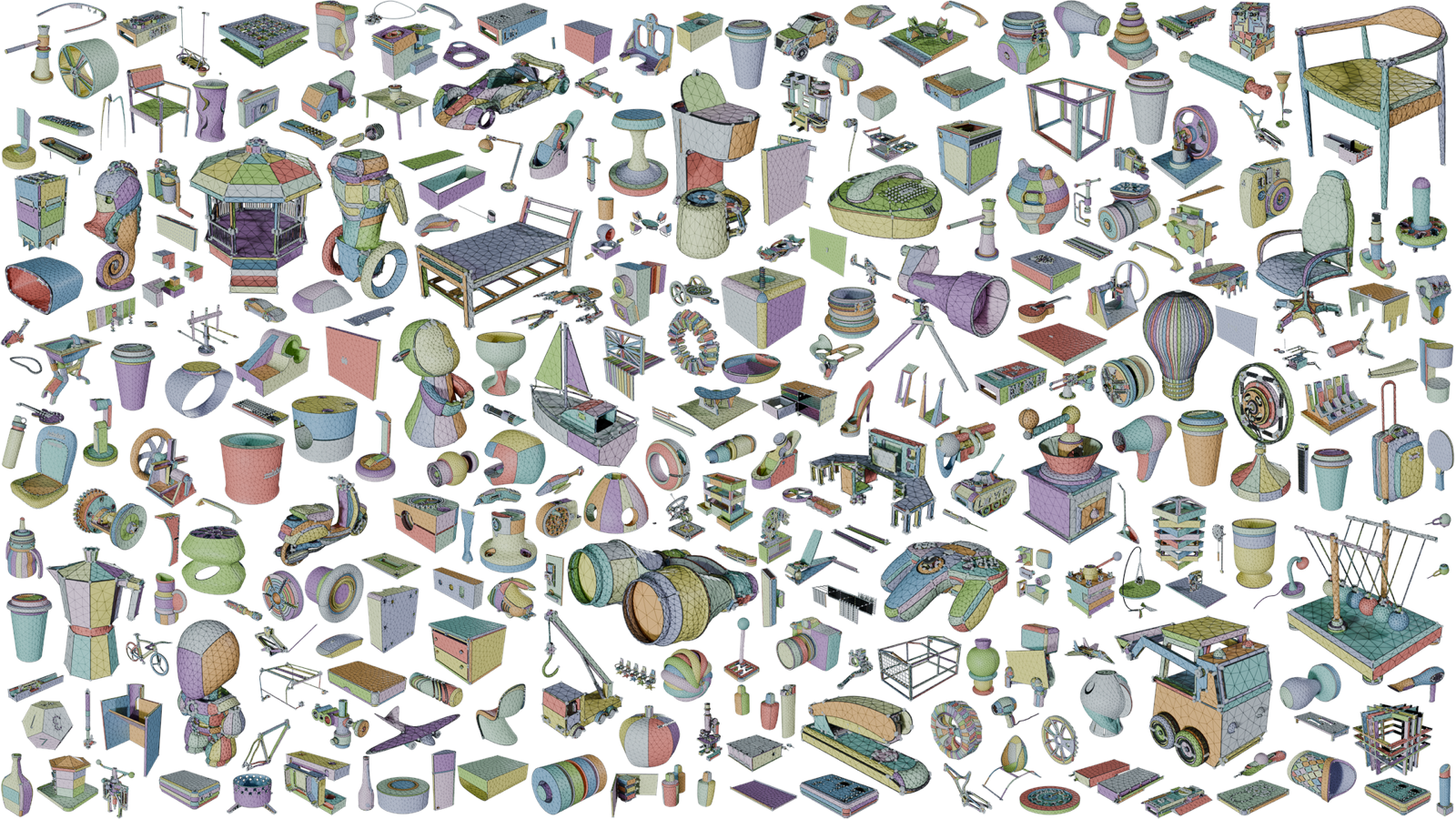}
  \vspace{-2.5em}
  \caption{
Selection of models from the ABC~\cite{Koch_2019_CVPR} and Fusion360~\cite{willis2021joinable} datasets meshed with our method; each B-Rep face is shown with a different color, illustrating preserved patch boundaries across complex models.
  }
\label{fig:teaser}
\end{teaserfigure}

%% file: 10-abstract.tex

Parametric boundary representation models (B‑Reps) are the de facto standard in CAD, graphics, and robotics, yet converting them into valid meshes remains fragile. The difficulty originates from the unavoidable approximation of high‑order surface and curve intersections to low‑order primitives: the resulting geometric realization often fails to respect the exact 
topology
encoded in the B‑Rep, producing meshes with incorrect or missing adjacencies. Existing meshing pipelines address these inconsistencies through heuristic feature-merging and repair strategies that offer no topological guarantees and frequently fail on complex models.

We propose a fundamentally different approach: the B‑Rep topology is treated as an invariant of the meshing process. Our algorithm enforces the exact B-Rep topology while allowing a single user‑defined tolerance to control the deviation of the mesh from the underlying parametric surfaces. Consequently, for any admissible tolerance, the output mesh is topologically correct; only its geometric fidelity degrades as the tolerance increases. This decoupling eliminates the need for post‑hoc repairs and yields robust meshes even when the underlying geometry is inconsistent or highly approximated.

We evaluate our method on thousands of real‑world CAD models from the ABC and Fusion 360 repositories, including instances that fail with standard meshing tools. The results demonstrate that topological guarantees at the algorithmic level enable reliable mesh generation suitable for downstream applications.

%% file: 11-intro.tex
\section{Introduction}



\begin{figure}
\includegraphics[width=\linewidth]{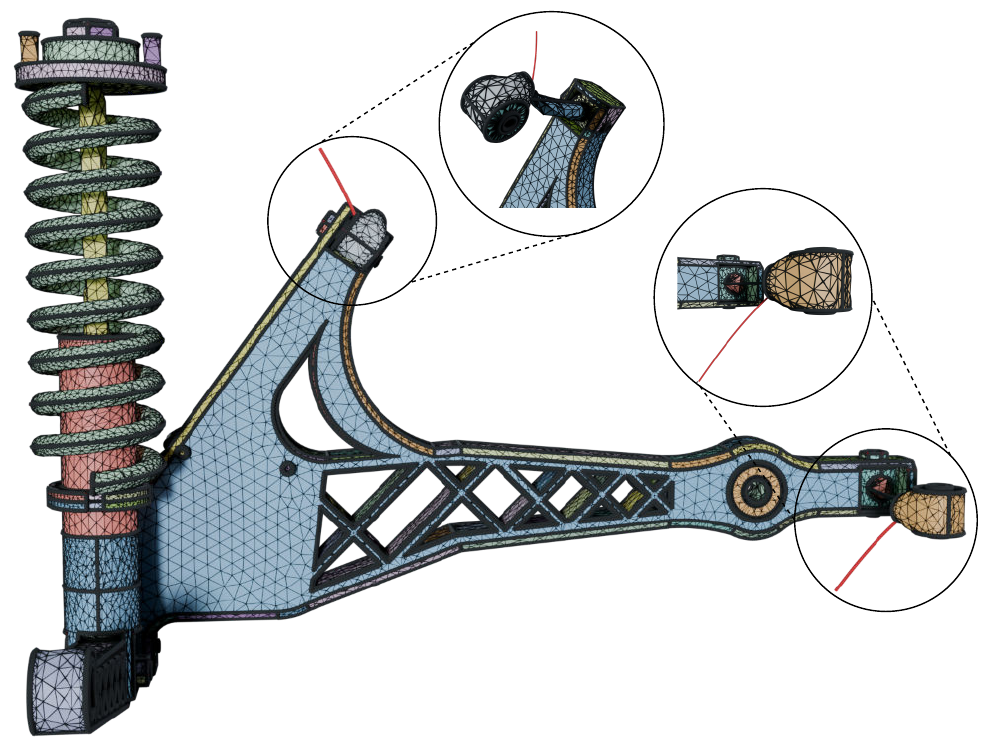}
\caption{Inconsistency between curves (black) and the 2D trimming curves lifted with the parameterization (red).
Since our method relies only on the 3D curves (black), it successfully generates a mesh despite these inconsistencies.} 
\label{fig:trim_vs_3D}
\end{figure}

Parametric models encoded as boundary representations (B-Reps) are the dominant representation in computer-aided design, manufacturing, and simulation.
A B-Rep describes a shape through parametric primitives (patches, curves, and points) to encode geometry together with a discrete combinatorial structure, i.e., the B-Rep's topology.
Patches are bounded by curves that are themselves bounded by points. The topology of a B-Rep is defined by an adjacency graph of such primitives.
While geometry and topology are tightly coupled in the representation, they play fundamentally different roles: topology encodes adjacency and incidence relationships, whereas geometry specifies how these entities are embedded in 3D as well as in parametric spaces. 
The geometry of curves is usually represented in two different forms:
first, their parametrized 3D geometry; second, they are projected onto the parametric domain of each adjacent patch, where they serve as trimming curves.
As a result, curves contain duplicated and potentially inconsistent geometric information (Figure~\ref{fig:trim_vs_3D}).
\begin{figure}
\includegraphics[width=\linewidth]{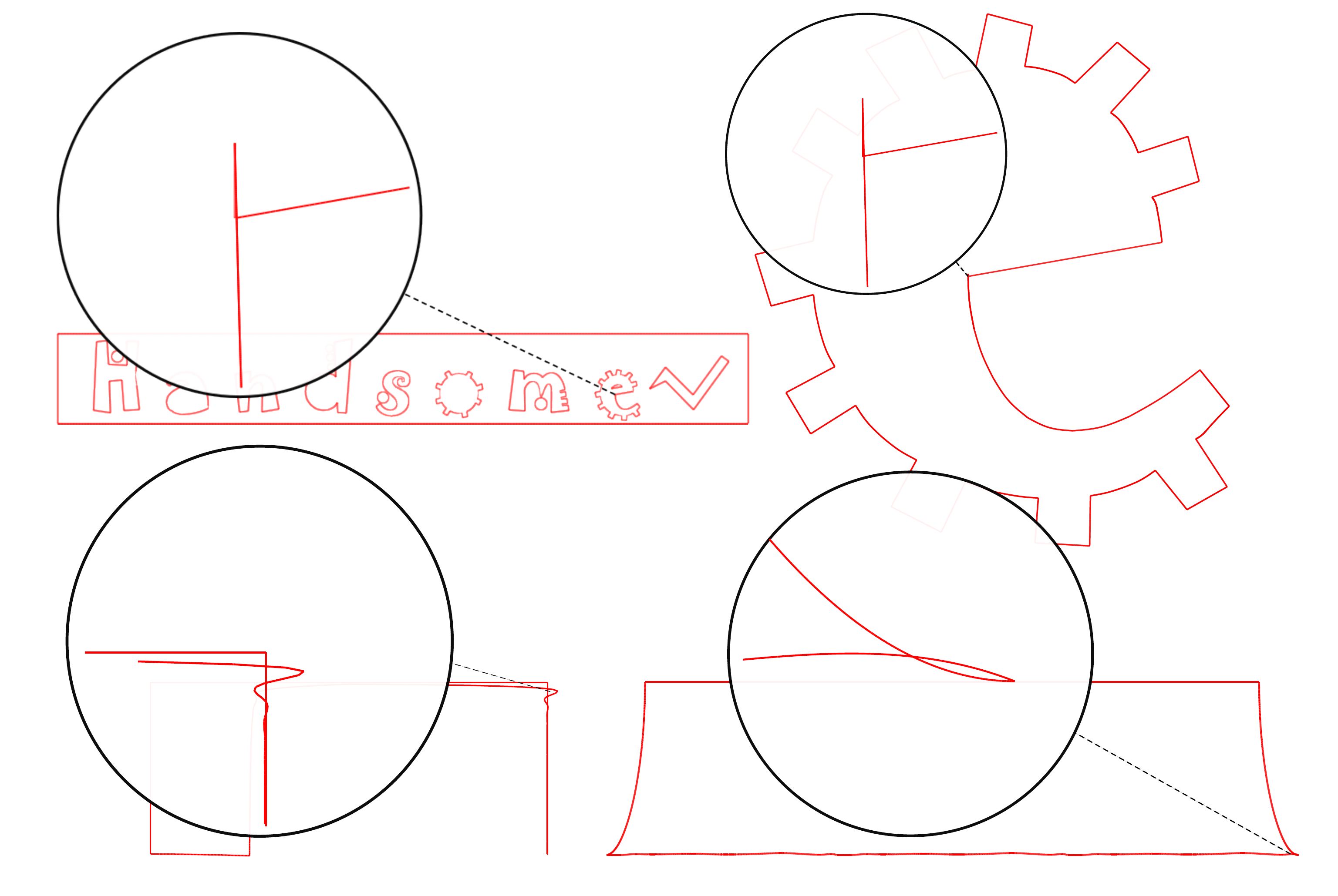}
\caption{Parametric domain of a patch with its 2D trimming curves. Although all curves are intended to form simple loops, approximation errors 
cause unintended intersections and artifacts.}
\label{fig:bad-param}
\end{figure}
The curve geometry in the parametric domain (traditionally called trims) is obtained from the 3D geometry by inverting the patch parameterization. This inversion requires solving a non-linear optimization problem and can introduce additional errors (Figure~\ref{fig:bad-param}); in our experiments, 14\% of trims in the Fusion dataset contain intersections\footnote{We detect intersections by densely sampling the trimming curves in the parametric domain.}.


Multiple downstream applications, such as finite element simulation or rendering, require converting B-reps to meshes.
However, this conversion remains a fragile and error-prone process.
The core difficulty arises from high-order parametric patch intersections and the need to approximate them with lower-order primitives.
In particular, exact intersections between parametric patches yield impractically high-order curves~\citep{KatzSederberg1988}, thereby forcing systems to rely on approximations (\Cref{sec:background}). 
As a result, the geometric realization of a B-Rep often becomes inconsistent with its topology (\Cref{fig:trim_vs_3D}).

Most existing meshing pipelines attempt to address these inconsistencies by employing geometric algorithms that merge features within an $\epsilon$-neighborhood or by applying heuristic repair strategies (\Cref{sec:related-work}). While effective for visualization and rendering, such approaches provide no guarantees on the correctness of the resulting topology and, depending on user parameters, fail to produce a mesh (\Cref{sec:comparison}). In practice, this results in undesired non-manifoldness, missing adjacencies, and costly downstream repairs, particularly for complex models.

In this work, we take a fundamentally different approach. Rather than resolving geometric inconsistencies heuristically, we treat the B-Rep topology as a discrete skeleton that is enforced throughout the meshing process. 
A key component is that we do not rely on the 2D trimming curves, as they are often flawed and may even self-intersect. Instead, we trim the parametric patches with their adjacent curves directly in 3D.
Building on this insight, we propose a meshing algorithm that \emph{guarantees preservation} of the B-Rep topology, independent of geometric approximations, while exposing a \emph{single} geometric parameter that controls the deviation between the mesh and the underlying parametric patches (\Cref{sec:method}). Note that our method is guaranteed to generate meshes with the correct topology, independent of the user-specified geometric tolerance; i.e., a larger tolerance yields a less geometrically accurate mesh, but has no influence on the topology. This decoupling of topological correctness from geometric approximation enables robust mesh generation even for B-Reps with inconsistent or badly approximated geometry.

We demonstrate the robustness of our method by successfully meshing over ten thousand B-Reps drawn from real-world CAD datasets (ABC~\cite{Koch_2019_CVPR} and Fusion 360~\cite{willis2021joinable}), including models that fail with standard meshing pipelines (\Cref{sec:comparison}). Our results show that enforcing topological guarantees at the algorithmic level eliminates the need for heuristic repairs and produces meshes suitable for downstream simulation and analysis.

%% file: 12-background-daniele.tex
\section{Preliminaries}
\label{sec:background}

\paragraph{CAD Kernels, Feature Trees, and BReps} A \emph{CAD kernel} is the geometric engine underlying most CAD systems: it implements the data structures and numerical algorithms used to \emph{create, edit, and query} solid models~\citep{Requicha1980,Mantyla1988}. Some popular kernels are 
Parasolid~\cite{siemens_parasolid},
ACIS/3D ACIS Modeler~\cite{spatial_acis_modeler},
CGM (Convergence Geometric Modeler)~\cite{spatial_cgm_modeler}, 
Granite (PTC Creo Granite Interoperability Kernel)~\cite{ptc_granite_kernel_pdf}, 
Autodesk ShapeManager (ASM)~\cite{autodesk_shapemanager_pressrelease}, 
Open CASCADE Technology (OCCT)~\cite{occt_dev_portal}, 
C3D Modeler~\cite{c3d_modeler}, and
SMLib~\cite{nvidia_smlib_intro}.
From the user's perspective, the \emph{input} to a CAD kernel is a construction history (often a \emph{feature tree}) that is, a sequence of operations such as extrude/revolve, fillet/chamfer, Boolean union/difference/intersection, shell/offset, or trim. 
The kernel evaluates this operation list and produces a final geometric model, typically stored as a \emph{boundary representation (B-Rep)}.

\paragraph{B-Reps} The \emph{output} B-Rep encodes both \emph{geometry} (parametric curves and surfaces, often NURBS)~\citep{PieglTiller1997} and \emph{topology} (how faces, edges, and vertices are connected)~\citep{Requicha1980,Mantyla1988}. 
A key detail is that B-Reps faces are rarely entire parametric surfaces: instead, each face is a \emph{trimmed patch}, i.e., a parametric surface restricted to a bounded subset of its $(u,v)$ domain by one or more trimming curves (loops) defined in parameter space.

\paragraph{Geometry and Topology Inconsistencies} Crucially, by construction, the geometry and topology of a B-Rep are \emph{not exactly consistent} except in trivial cases. 
Many kernel operations require computing surface--surface intersections and projecting curves between 3D space and multiple parameter domains; the exact intersection curves can have extremely high algebraic complexity even for low-degree surfaces~\citep{KATZ1988253}. 
Practical kernels, therefore, rely on approximation, subdivision, and tolerances. 
As a result, the same ``logical'' entity (e.g., a shared edge) may have multiple representations, a 3D curve plus separate trimming curves on each adjacent face, that cannot be made to coincide everywhere, leading to small gaps, overlaps, or mismatched endpoints \Cref{fig:bad-param}.

\paragraph{Meshing}
This inconsistency is one reason a B-Rep is often \emph{not directly usable} for downstream applications such as rendering or simulation, which typically require a discrete mesh. 
Extracting a valid surface or volume mesh from a B-Rep is challenging precisely because meshing must reconcile these geometric/topological mismatches: tiny gaps can create cracks, overlaps can introduce self-intersections, and local inconsistencies can break watertightness or manifoldness. 
Robust meshing, therefore, cannot be treated as a routine discretization step; it must explicitly handle the inherent inconsistencies of real CAD B-Reps.

\paragraph{Our Contribution} We propose a novel method to convert B-Reps into meshes that is unconditionally robust to these inconsistencies: our key observation is to avoid using 2D trimming curves entirely and instead reconstruct them by tracing directly in 3D over patches using closest point projection paired with a topologically robust tracing algorithm. Our contribution is orthogonal to the algorithms within a CAD kernel, and our algorithm can process B-Reps produced by any CAD kernel.

%% file: 20-related.tex
\section{Related Work}
\label{sec:related-work}

B-reps come from open-source CAD kernels, like OpenCascade~\citep{occt} or closed-source CAD kernels,like Parasolid and ASIC, but the inherent inconsistencies of B-reps always emerge.
Therefore, even if we can mesh each parametric patch~\citep{Rockwood1989,Cripps2011}, constructing meshes with triangulated patches that are connected according to the patch topology is difficult.
In this section, we discuss several approaches others have used to convert the parametric models generated by these kernels into triangulated surfaces.

\paragraph{Top-down methods} One class of approaches for constructing a triangle mesh from a parametric model typically involves triangulating each parametric patch and then stitching them together according to the topology of the parametric model~\cite{CUILLIERE1998139,Sheng1992}.
The boundaries of these per-patch triangulations depend on the parametric coordinates of trims, which can self-intersect (Figure~\ref{fig:bad-param})
and therefore cannot guarantee aligned boundaries between patches, resulting in a difficult stitching process~\citep{Kahlesz2002,barequet1997repairing}.
Some of these methods do not require knowledge of the input model's topology, such as \citep{wei2024scalablefieldalignedreparameterizationtrimmed}, which uses local matching processes to remove gaps after an initial stitching step produces invalid results.
As they also rely on the trimmed patches, they are vulnerable to inconsistencies in the trimming curves~(\Cref{fig:trim_vs_3D}).

\paragraph{Repair} Stitching per-patch meshes involves ``repairing'' the mesh between trims, and advances have been made to repair meshes that come from parametric models.
Some methods, such as \citet{Guo2019}, improve the per-surface meshes created by CAD kernels to ease the stitching process, while others, such as \citet{Wen_2025}, perform feature detection while repairing the full meshes generated by CAD kernels.
Some have even designed user interfaces to let users manually repair gaps~\citep{Zheng2001}, while others suggest repairing problematic regions by using voxel grids and
implicit surface meshing to replace problematic regions~\citep{bischoff2005structure,ju2004robust}, or relying on user-tolerance to merge nearby patches \cite{Busaryev2009}.
Such techniques enjoy great generality, as they are applicable to triangle meshes, but because they lack the input topology, they cannot guarantee that the meshes they generate follow the original parametric model's geometry.
Rather than relying on repairing inconsistencies introduced by parametric trims, our method guarantees a consistent trim that respects the topology of the input model by tracing a 3D trim directly into each patch.



\paragraph{Bottom-up methods} The open source softwares GMsh~\citep{Geuzaine2009} and OCCT~\citep{occt}, and the method from \citet{Li2025} implement bottom-up approaches for triangulating or quadrangulating~\citep{Reberol2021} B-Reps.
That is, they linearize the boundary curves between patches and then triangulate each patch to match those linearized boundaries, trivially guaranteeing that the boundaries between curves conform to one another.
This triangulation process requires using the parametric trims as the boundary of a $2D$ triangulation algorithm like constrained Delaunay triangulation~\citep{Shewchuk1996,PaulChew1989} or advancing front methods~\citep{Lo1985,Peraire1987, Liu_2024}, but because the parametric trim curves are truncated these boundaries can be difficult to mesh due to unfortunate features like self-intersections or poor sampling~(\Cref{fig:bad-param}), resulting in challenges during triangulation.
Although our approach prioritizes data from the bottom up, we mesh our patches and curves in 3D to ensure they are sufficiently sampled, and then perform trimming in 3D, thereby sidestepping the issue of inconsistent trimming curves in the patches' parametric domains.

\paragraph{Adaptive trimming} 
Rather than putting effort into matching patches and trims, some have attempted to change the trims themselves.
For example, \citet{Wang2025} propose their own method for determining trims and optimizing the resulting curves, whereas~\citet{Jourdes2014} utilizes raycasting to re-generate trims by projecting the trimmed boundaries of surface $A$ upon surface $B$ to generate a new parametric curve on $B$ before meshing the re-trimmed mesh of $B$ to guarantee a conforming mesh.
More globally, aligned curve meshing~\citep{Yang2025,Yu2021,Yang_2025} employs Delaunay meshing on the edge network~\citep{Xiao2021} to mesh patch boundaries. 
This ensures that edges do not intersect in 3D and that the network of meshed trims is topologically correct before meshing each surface.
These methods still rely on constrained triangulations using trim curves in the parametric space, whereas we use only the 3D coordinates of our edges to avoid inconsistent representations across different meshes.

%% file: 30-method.tex
\section{Method}
\label{sec:method}

\begin{figure}
    \centering\footnotesize
    \includegraphics[width=\linewidth]{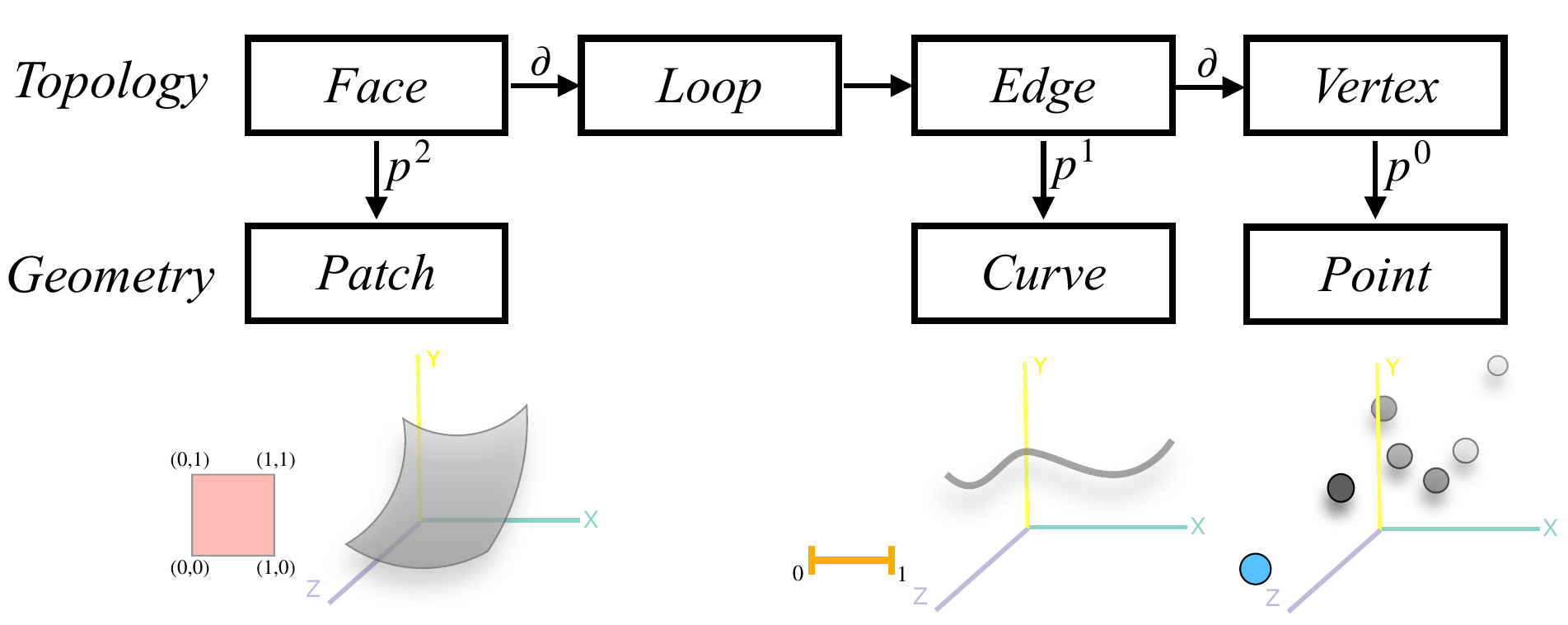}\hfill
    \caption{A B-Rep contains topology, a combinatorial structure in which faces are bounded by loops of edges and edges by vertices, and geometry, which embeds these entities as 3D surfaces, curves, and points.}
    \label{fig:brep}
\end{figure}

The input to our algorithm is a boundary representation (B-Rep), consisting of a topology together with an associated geometric embedding (\Cref{fig:brep}).
In our implementation, we use the ABS dataset~\cite{abs}, which provides B-Reps converted from STEP files. Note that models in the ABC and Fusion datasets are generated using different CAD kernels.

\begin{figure}
\centering\footnotesize
         \includegraphics[width=\linewidth]{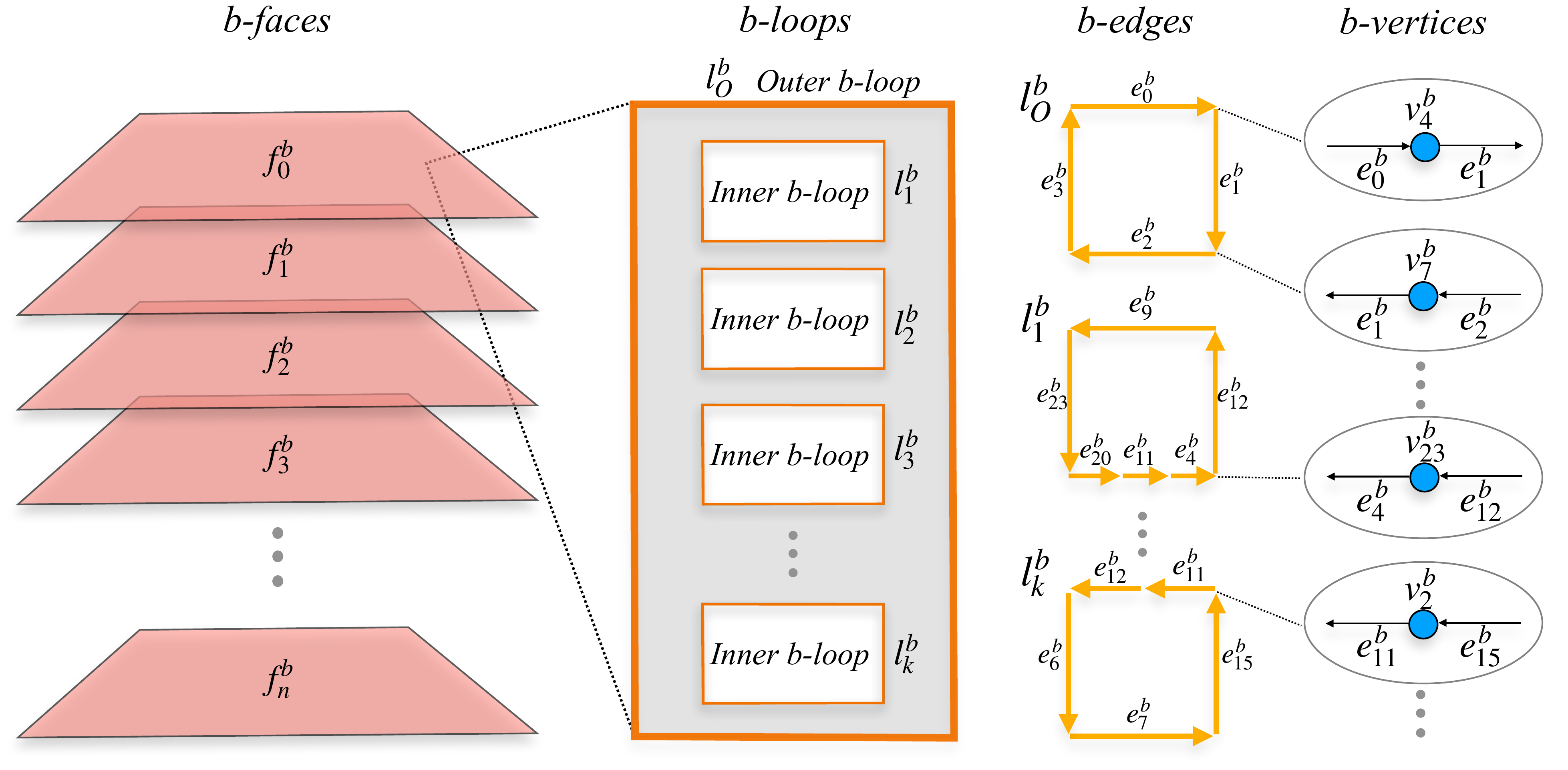}
\caption{Overview of the topological entities and their relationships.}
\label{fig:topo}
\end{figure}

\paragraph{Topology}
The topology $\mathcal{P}$ of a B-Rep is composed of a set of vertices, edges, loops, and faces (Figure~\ref{fig:topo}).  
To distinguish them from the mesh primitives (such as triangle faces and vertices) that appear in our output later, we refer to them as  \emph{b-vertices} $\{v^b_i\}$, \emph{b-edges} $\{e^b_i\}$, \emph{b-loops} $\{\ell^b_i\}$, and \emph{b-faces} $\{f^b_i\}$.

\begin{figure}
\centering\footnotesize
         \includegraphics[width=\linewidth]{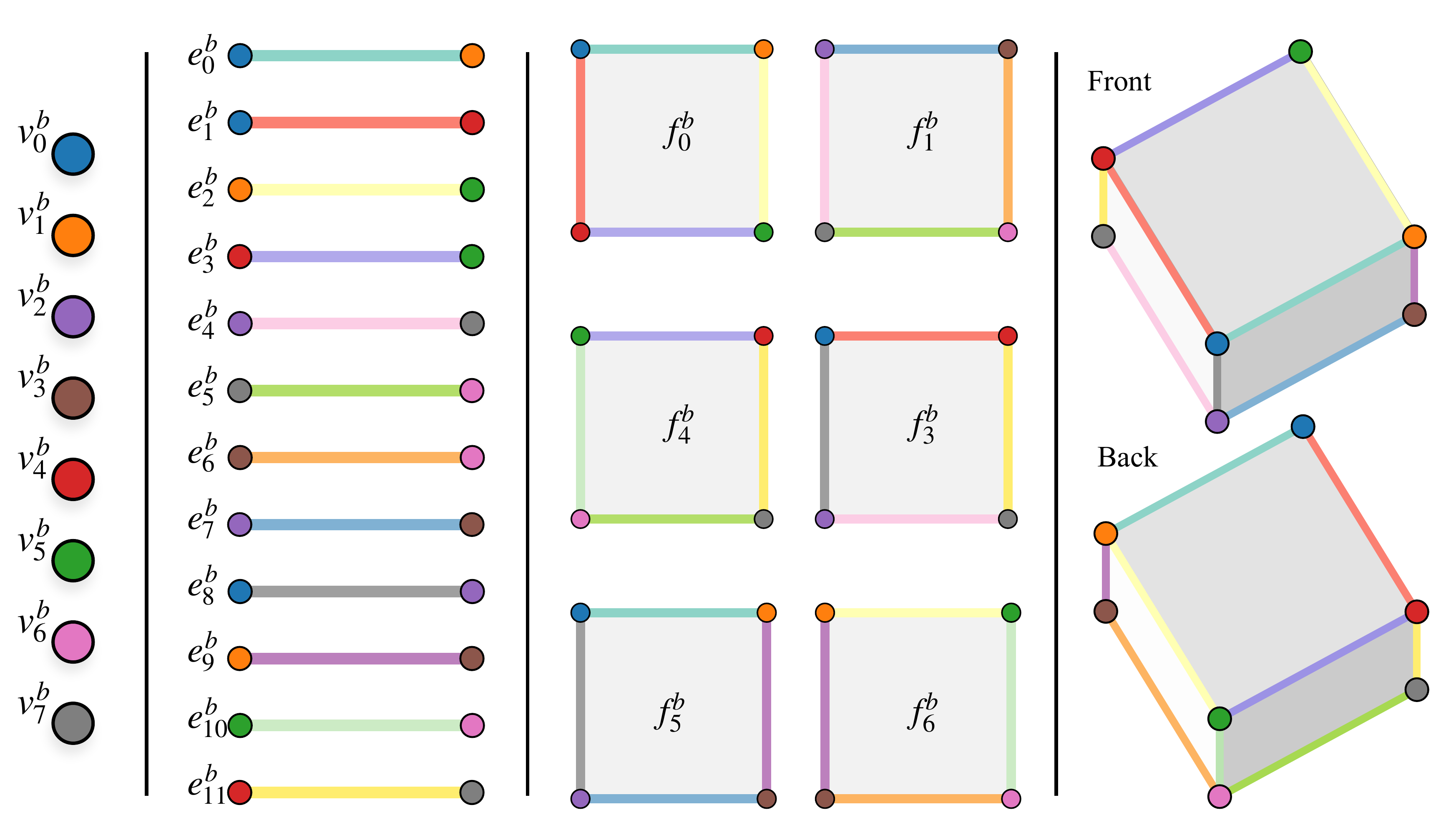}\hfill
\caption{B-Rep topology of a cube. The same b-vertices and b-edges are shown with the same color, highlighting the correspondence between b-vertices, b-edges, b-loops, and b-faces.}
\label{fig:cube_topo}
\end{figure}

A \emph{b-edge} is an oriented topological entity defined by an ordered pair of b-vertices $(v^b_1, v^b_2)$ and incident to one or more b-faces. A \emph{b-face} has a unique outer b-loop $\ell^b_O$ and may contain a collection of $k$ inner b-loops $(\ell^b_1, \ldots, \ell^b_k)$ representing holes. Each \emph{b-loop} is a closed sequence of oriented b-edges $(e^b_1, \ldots, e^b_n)$. 
For instance, a cube has 8 b-vertices, 12 b-edges, and 6 b-faces, all connected to obtain the correct topology (\Cref{fig:cube_topo}).

\begin{figure}
\centering
\includegraphics[width=\linewidth]{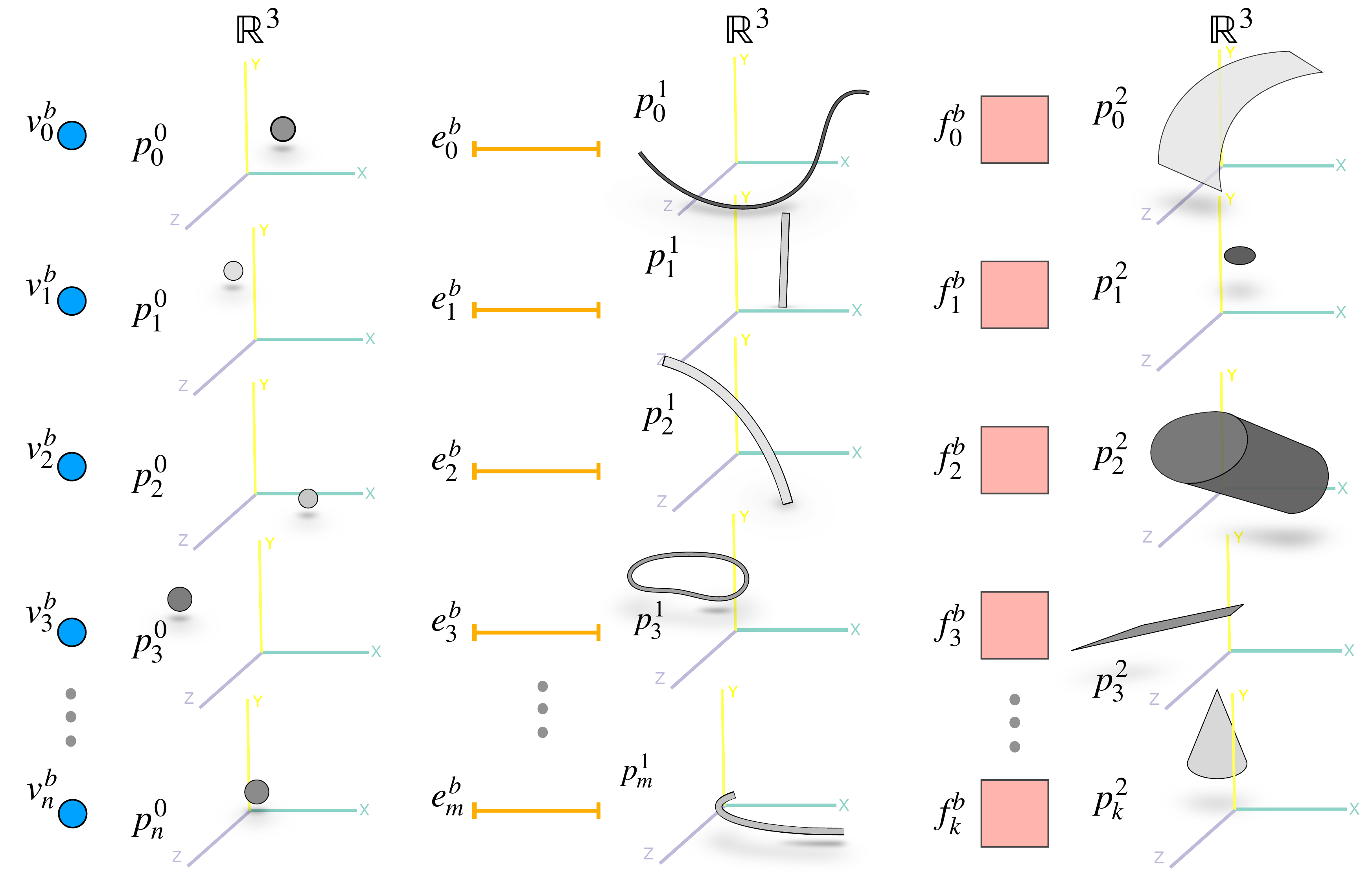}
\caption{Overview of the geometrical entities.}
\label{fig:geometry}
\end{figure}

\paragraph{Geometry}  
The topological entities have their corresponding parametric shape to embed them in $\mathbb{R}^3$  (Figure~\ref{fig:geometry}).  
\begin{itemize}
    \item Each b-vertex $v^b_i$ has a corresponding point $p^0_i \in \mathbb{R}^3$.  
    \item Each b-edge $e^b_i$ has a parametric curve $p^1_i(t) : [0,1] \to \mathbb{R}^3$ mapping the unit interval to a 3D curve.  
    \item Each b-face $f^b_i$ has a parametric surface $p^2_i(u,v)z : [0,1] \times [0,1] \to \mathbb{R}^3$ mapping the unit square $[0,1]^2$ to a 3D surface.  
\end{itemize}
Note that we do not assume anything about the geometry of the topological entities. For instance, the 3D curve $p^1_i(t)$ attached to the b-edge $e^b_i$ can be arbitrarily far from the parametric surfaces attached to the b-faces adjacent to $e^b_i$. In particular, the geometric embedding may be arbitrarily inconsistent with the topology (\Cref{fig:trim_vs_3D}), reflecting the need for tolerance-driven approximations present in other methods (\Cref{sec:related-work}).

\paragraph{Output} 
The output of our algorithm is a triangle mesh with the same topology as $\mathcal{P}$.
A single geometric tolerance parameter $\epsilon$ controls the approximation error between the mesh and the input parametric geometry, without affecting topological correctness.
We emphasize that, regardless of the input geometry, our meshing algorithm \emph{preserves the input topology exactly}. In other words, the hierarchy of faces, loops, edges, and vertices, together with their structural relations, is reproduced \emph{identically} in the mesh output.

\subsection{Algorithm}\label{sec:algo}

\begin{figure*}
    \centering\footnotesize
    \includegraphics[width=\linewidth]{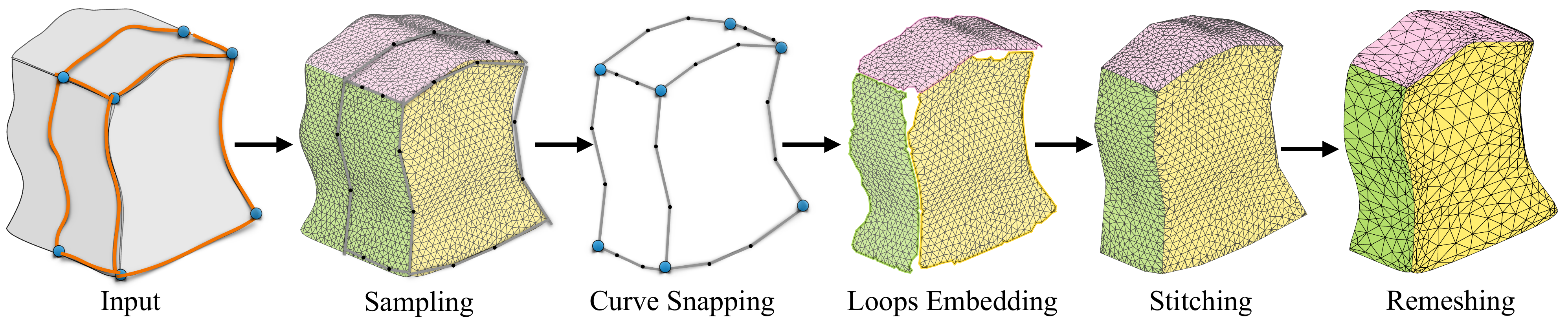}
\caption{Overview of our five-stage algorithm: we start by sampling every primitive, snapping curves to vertices, and embedding them on face meshes. We then stitch the trimmed meshes to obtain a topologically valid surface mesh, which we finally remesh to improve element quality.}
    \label{fig:pipeline}
\end{figure*}

Given a B-Rep, we first construct independent polylines and triangle meshes for each parametric curve and surface.
We snap polyline endpoints to vertices and directly embed each polyline into its corresponding 3D surface mesh, thereby trimming the mesh.
We then stitch the resulting trimmed surface meshes along shared b-edges to recover the full model, and finally remesh it to improve element quality while preserving topology (\Cref{fig:pipeline}).
Our algorithm proceeds in five stages: (1) geometry sampling (Section~\ref{sec:sampling}), (2) curve snapping (Section~\ref{sec:curve-snap}),(3) loop embedding on each face (Section~\ref{sec:tracing}), (4) stitching across faces (Section~\ref{sec:stich}), and (5) geometry- and topology-preserving isotropic remeshing (Section~\ref{sec:isotropic}).
Where applicable, we employ several heuristic strategies to improve performance and geometric accuracy without affecting correctness (Section~\ref{sec:acceleration}).

\paragraph{Correctness}
The base algorithm guarantees preservation of the input B-Rep topology by construction. 
The loop embedding stage embeds each b-loop as a simple cycle that partitions the mesh according to the known topology of the loop. 
The stitching stage preserves this topology by aligning boundary discretizations without introducing or removing adjacency relations.
Any heuristic acceleration used in the pipeline is validated using purely topological criteria (rather than approximate geometric ones) and reverts to the base algorithm if validation fails.


\subsection{Stage 1. Sampling}\label{sec:sampling}
The goal of the sampling stage is to construct a sufficiently refined simplicial approximation of each geometric primitive of our input B-Rep.
This stage is purely geometric and does not encode any topological decisions.

\paragraph{b-vertices} We do not need to sample the b-vertices as we already have their corresponding 3D points. 

\paragraph{b-edges} We sample every curve $p^1_i$ with a polyline consisting of vertices $v^e_i$ and edges $e^e_i$. Starting from a single edge, we recursively subdivide it until the distance of the polyline to the curve is less than $\epsilon$.

\paragraph{b-faces}
For the patch associated with a b-face $f^b_i$, we mesh its 2D parametric domain using MMG~\cite{balarac2022tetrahedral} with the patch’s induced metric, and lift the resulting triangulation to 3D via the parametric surface, obtaining the mesh $M_i$. Note that we mesh the entire parametric domain and ignore any trimming at this stage.
Furthermore, we perform RGB subdivision~\cite{puppo2009rgb} until the per-triangle distance between the mesh and its corresponding patch is less than $\epsilon$ and the maximum edge length is below 0.01\% of the diagonal of the model.
Since computing the exact distance to a parametric surface is difficult, we conservatively estimate.
We sample quadrature points in each parametric triangle (order 5), lift them both using the linear (barycentric) interpolation of the mesh triangle and the parametric surface evaluation, and integrate the distance between the resulting 3D points.


\subsection{Stage 2. Curve Snapping}\label{sec:curve-snap}
After sampling all geometric primitives, we snap the endpoints of each sampled b-edge curve to the points corresponding to their incident b-vertices. 
Since the sampled endpoints generally do not coincide exactly with the b-vertex positions, snapping inevitably introduces geometric error.
To avoid localized distortion, we distribute this error along the curve by applying a 1D uniform Laplacian smoothing to displacements.
This step enforces consistency between the sampled curves and the B-Rep topology and ensures that curve endpoints coincide exactly with the embedded b-vertices before loop embedding.


\subsection{Stage 3. Loop Embedding}\label{sec:tracing}
This stage is performed independently for each b-face $f^b_k$.
The goal is to embed all b-loops $\ell^b_i$ belonging to $f^b_k$, together with their b-vertices $v^b_j$, onto the mesh $M_k$ of the corresponding parametric surface (Algorithm~\ref{alg:trace-main}).
Specifically, we create an edge chain on the triangle mesh $M_k$ corresponding to each b-edge in each b-loop $\ell^b_i$.
During this process, we maintain correspondence between mesh edges and b-edges, as well as the association between embedded vertices and their originating b-vertices.

\begin{algorithm}
\caption{Loop embedding for a b-face $f^b_k$}
\label{alg:trace-main}
\begin{algorithmic}[1]
\Require Mesh $M_k$ sampled from the parametric patch of $f^b_k$; b-loops $\{\ell^b_O,\ell^b_1,\ldots,\ell^b_k\}$ with sampled polylines
\Ensure Mesh $M_k$ with all loops embedded as mesh edge chains

\Statex
\State \textbf{Preprocess non-manifold loops}\label{ll:nnm}
\State Duplicate every non-manifold b-loop edge and vertex
\State Record a merge map for duplicated entities

\Statex
\State \textbf{Embed outer loop}\label{ll:outer}
\State $M_k \gets \Call{EmbedLoop}{M_k,\ \ell^b_O,\ \textsc{KeepDisk}}$

\Statex
\State \textbf{Embed inner loops}
\For{$r \gets 1$ to $k$}
    \State $M_k \gets \Call{RestoreDisk}{M_k}$
    \State $M_k \gets \Call{EmbedLoop}{M_k,\ \ell^b_r,\ \textsc{KeepComplement}}$
\EndFor

\Statex
\State \textbf{Restore original topology}
\State Merge duplicated vertices and edges by collapsing recorded chains

\State \Return $M_k$
\end{algorithmic}
\end{algorithm}

The core idea of our algorithm is to embed the boundary of a disk onto a disk, without relying on geometric heuristics.
The parametric patch and its associated mesh $M_k$ are topological disks; however, the input b-loops may be non-manifold and therefore may not enclose disks.
To address this, we temporarily duplicate every non-manifold b-loop edge and vertex, while keeping track of the duplication (Algorithm~\ref{alg:trace-main}, line~\ref{ll:nnm}).
This operation yields a set of manifold b-loops, each of which encloses a topological disk, without altering the represented topology.
We then embed each such disk boundary onto the disk $M_k$.

\begin{algorithm}
\caption{\textsc{EmbedLoop}}\label{alg:embed-loop}
\begin{algorithmic}[1]
\Require Mesh $M$ (topological disk); b-loop $\ell=\{e^b_i\}_{i=1}^n$
\Ensure Mesh $M$ with $\ell$ embedded as a simple cycle

\State Initialize forbidden edge set $\mathcal{F} \gets \partial M$
\State Initialize embedded cycle $\Gamma \gets \emptyset$

\For{$i \gets 1$ to $n$}
    \State Let $e^b_i=(v^b_a,v^b_b)$ with sampled points $\{v^e_{i,j}\}_{j=1}^m$
    \State Project $v^b_a$ onto $M$ and denote the mesh vertex by $s$\label{ll:start}
    \For{$j \gets 2$ to $m$}
        \State Project $v^e_{i,j}$ into $M$; call it $e$\label{ll:second}
        \State $\pi \gets$ shortest path from $s$ to $e$ on the edge graph of $M$ avoiding $\mathcal{F}$\label{ll:dij}
        \State $\Gamma \gets \Gamma \cup \pi$
        \State $\mathcal{F} \gets \mathcal{F} \cup \pi$\label{ll:avoid}
        \State Enforce a simplicial embedding of $\mathcal{F}$ by local refinement~\cite{zint2025topological}
        \State $s \gets e$
    \EndFor
\EndFor

\State Cut $M$ along $\Gamma$, yielding components $(M_1,M_2)$
\If{\textsc{KeepDisk}}
    \State $M \gets$ disk component among $\{M_1,M_2\}$\label{ll:kdisk}
\Else
    \State $M \gets$ complementary component\label{ll:ddisk}
\EndIf

\State \Return $M$
\end{algorithmic}
\end{algorithm}

We begin by tracing the outer loop $\ell^b_O=\{e_i^b\}, i=1,\dots,n$ made by the $n$ b-edges $e_i^b$ (Algorithm~\ref{alg:trace-main}, line~\ref{ll:outer}).
We select the first b-edge $e^b_1 = (v^b_j, v^b_k)$ and project the point $p^0_j$ associated with the b-vertex $v^b_j$ onto the closest triangle on the mesh $M_k$ (Algorithm~\ref{alg:embed-loop} line~\ref{ll:start}); we denote the resulting mesh vertex by $s$. 
Let $\{ v^e_{1,i} \}_{i=1}^{m}$ denote the sampled points of the parametric curve associated with $e^b_1$.
We insert the second sampled point $v^e_{1,2}$ into $M_k$ by splitting the closest triangle to $v^e_{1,2}$ on $M_k$  (Algorithm~\ref{alg:embed-loop} line~\ref{ll:second}); we call this point $e$.

\begin{figure}
    \centering
    \includegraphics[width=\linewidth]{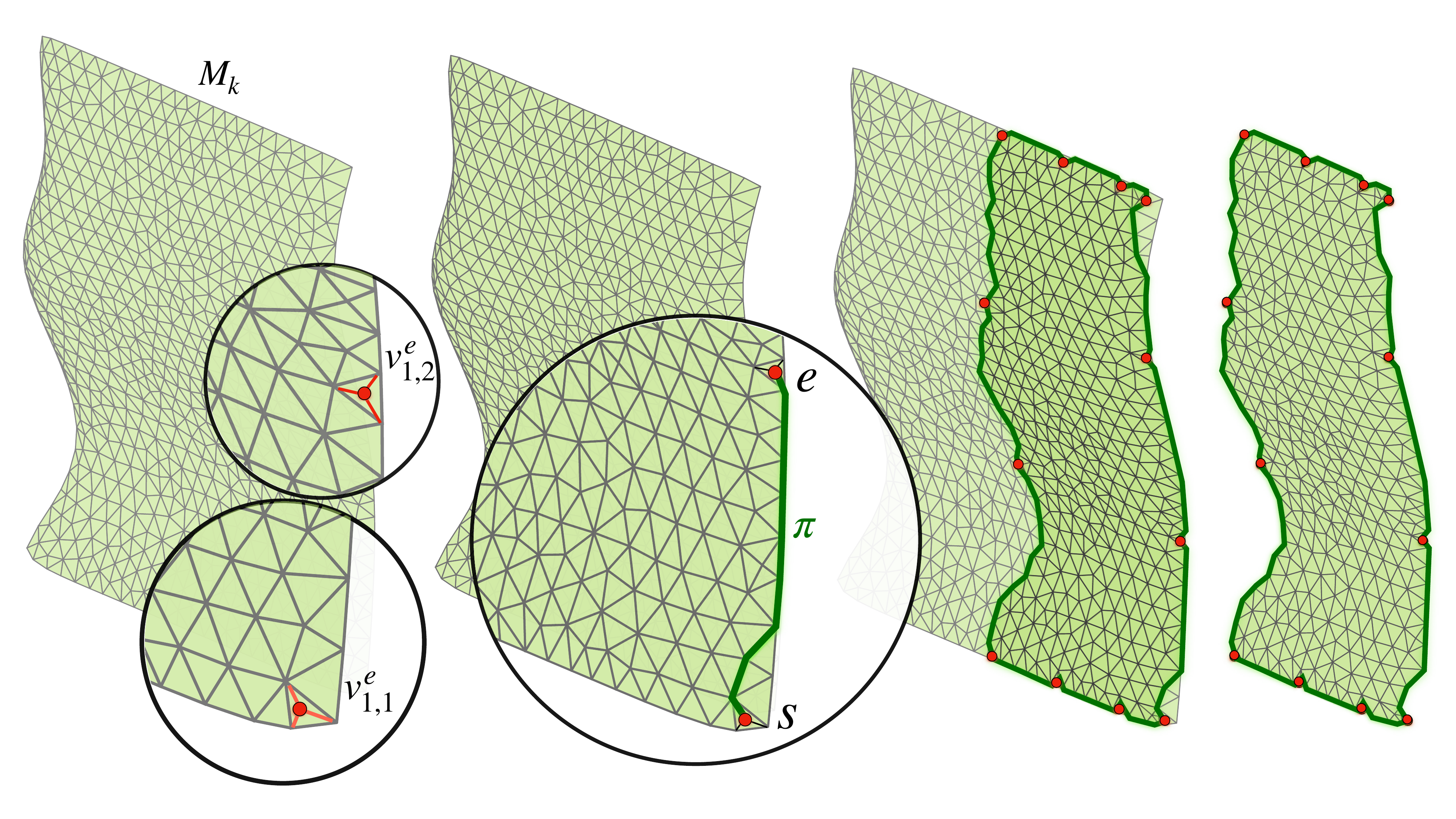}
\caption{Overview of our tracing algorithm: we start by projecting two points, trace an edge chain on the mesh using Dijkstra’s algorithm, and repeat this process for each successive vertex.}
    \label{fig:tracing}
\end{figure}

To embed the first sampled segment $\{ v^e_{1,1}, v^e_{1,2}\}$ of $e^b_1$ on the mesh, we compute a shortest path $\pi$ between the two embedded vertices $s$ and $e$ on $M_k$ using Dijkstra's algorithm over the edges of $M_k$ (Algorithm~\ref{alg:embed-loop} line~\ref{ll:dij}, \Cref{fig:tracing}).
To preserve the correct disk topology, boundary and previously traced edges are excluded from the graph traversal  (Algorithm~\ref{alg:embed-loop} line~\ref{ll:avoid}).
This restriction prevents the embedded path from crossing existing boundaries and ensures consistency with the intended topological partition.
We then repeat the same procedure with the next vertex $v^e_{1,2}$ until $\ell^b_O$ is fully embedded in $M_k$. 
Note that if the mesh is too coarse, there may not be a path between the two vertices.  By making the mesh a simplicial embedding~\cite{zint2025topological} of all previously traced edges, we guarantee that Dijkstra's algorithm always has a solution. 
Once the outer loop has been fully traced, the embedded simple cycle partitions $M_k$ into two connected components: a disk and an annulus. Since the outer b-loop is known a priori to enclose a disk, this decision is purely topological and does not rely on geometric predicates. We therefore retain the disk component and discard the annulus (Algorithm~\ref{alg:embed-loop} line~\ref{ll:kdisk}).

\paragraph{Inner-loop tracing}
After embedding the outer loop, the resulting mesh remains topologically a disk and serves as the input for tracing the inner b-loops (holes) using the same procedure.
Unlike the outer loop, for inner loops we discard the disk component and retain the complementary component (Algorithm~\ref{alg:embed-loop}, line~\ref{ll:ddisk}).
Once the second inner loop is traced, the mesh is no longer a disk but an annulus.
To restore the disk property required for subsequent embeddings, we select one vertex on each boundary component and trace a connecting path between them~\cite{cuttingdisk}.
This operation is purely topological and restores the disk property without affecting the final connectivity of the mesh.

After all loops have been embedded, we restore the original B-Rep topology, which was temporarily altered to ensure manifold b-loops during embedding, by undoing the preprocessing duplication by merging duplicated entities.
Duplicated b-edges are re-identified by stitching the corresponding boundary edge chains, as described in \Cref{sec:stich}.
Duplicated b-vertices are restored by collapsing the chains of mesh edges connecting their copies.
This operation is the inverse of the initial duplication step and restores the correct B-Rep topology.


\subsection{Stage 4. Stitching}\label{sec:stich}

\begin{figure}
    \centering
    \includegraphics[width=0.8\linewidth]{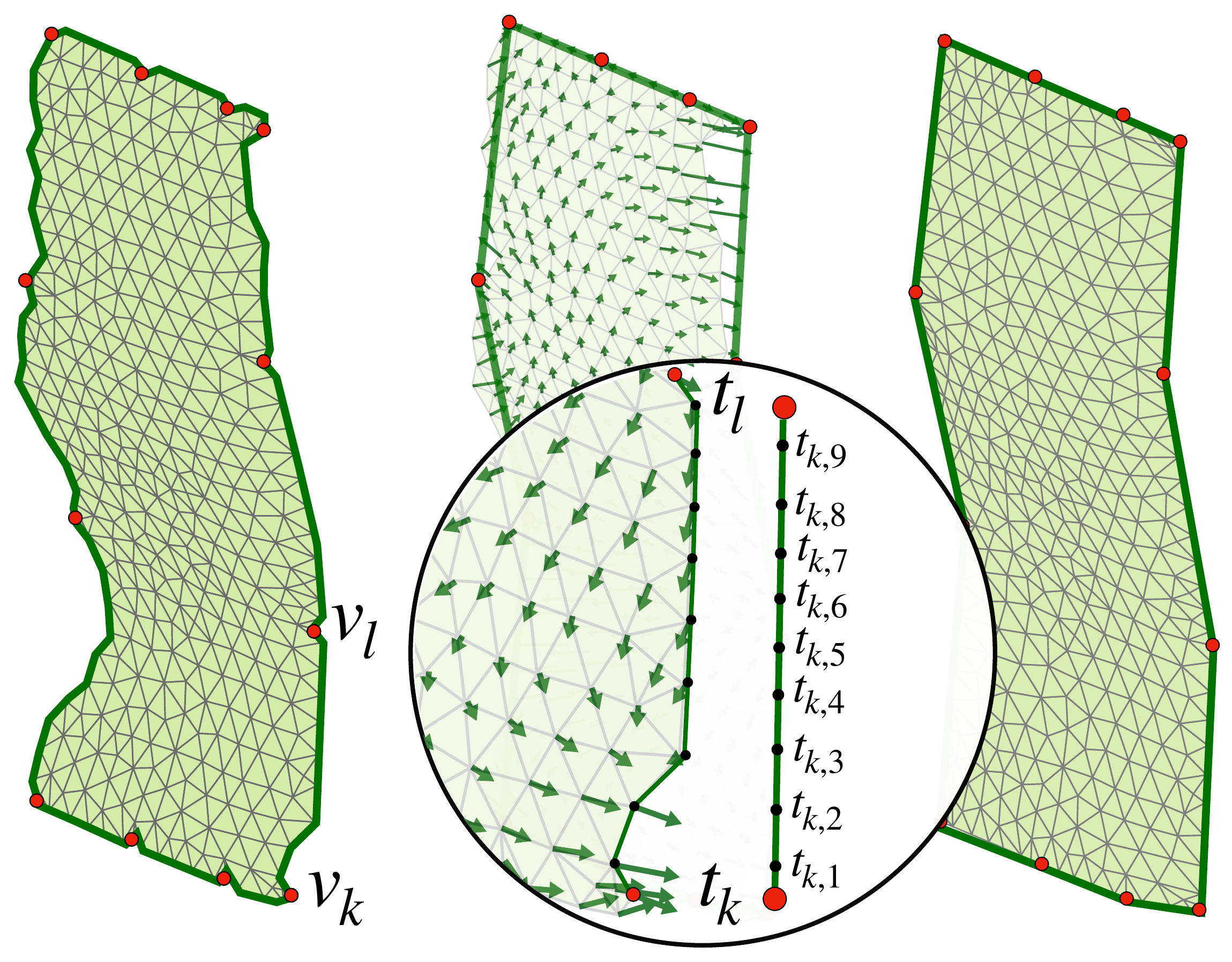}
     \caption{Overview of snapping: after we embed the b-edge $(v_k, v_l)$ onto the surface mesh, we snap its vertices to the target curve, and distribute the resulting snapping displacement over the mesh (green arrows), leading to a surface that conforms to the b-edge geometry.}
    \label{fig:stich-ex}
\end{figure}

The result of the previous stage is a collection of trimmed meshes, one per b-face, whose boundaries are topologically consistent but discretized with different numbers of vertices.
Additionally, every boundary edge $e_e^j$ is associated with an input b-edge $e^b_j$, and its endpoints have a corresponding parametric value. 
That is, for the edge $e_e^j=\{v_k, v_l\}$, the two vertices $v_k$ and $v_l$ have the associated parameters $t_k$ and $t_l$ on the curve attached to the b-edge $e^b_j$.
If $v_k$ is not the point of a b-vertex, $t_k$ is the parametric value of the sampled point on the curve; if $v_k$ was traced with Dijkstra's algorithm, it is arc-length interpolated from the two projected points to reduce distortion.

To generate the final mesh, we snap every mesh vertex corresponding to a b-vertex to its associated b-vertex position,
and every boundary vertex to the position described by its associated curve and parameter $t_i$. Since the points and curves are not necessarily on the mesh, the snapping introduces a localized geometric discrepancy near the boundary (\Cref{fig:stich-ex}).
To distribute this error, we smooth it over the mesh by computing the boundary displacement $u_c$ and solving for
\[
\Delta u = 0\quad\text{with}\quad u\rvert_{\partial M_k}=u_c,
\]
with $\partial M_k$ the boundary of $M_k$ and $\Delta$ the uniform Laplace operator, as cotangent weights may become unstable on low-quality or highly anisotropic meshes.
Note that triangles may contain two vertices on $\partial M_k$ without the corresponding edge lying on the boundary; therefore, before assembling and solving the Laplace equation, we ensure that $M_k$ is a simplicial embedding of its boundary $\partial M_k$.
\begin{figure}
    \centering
    \includegraphics[width=\linewidth]{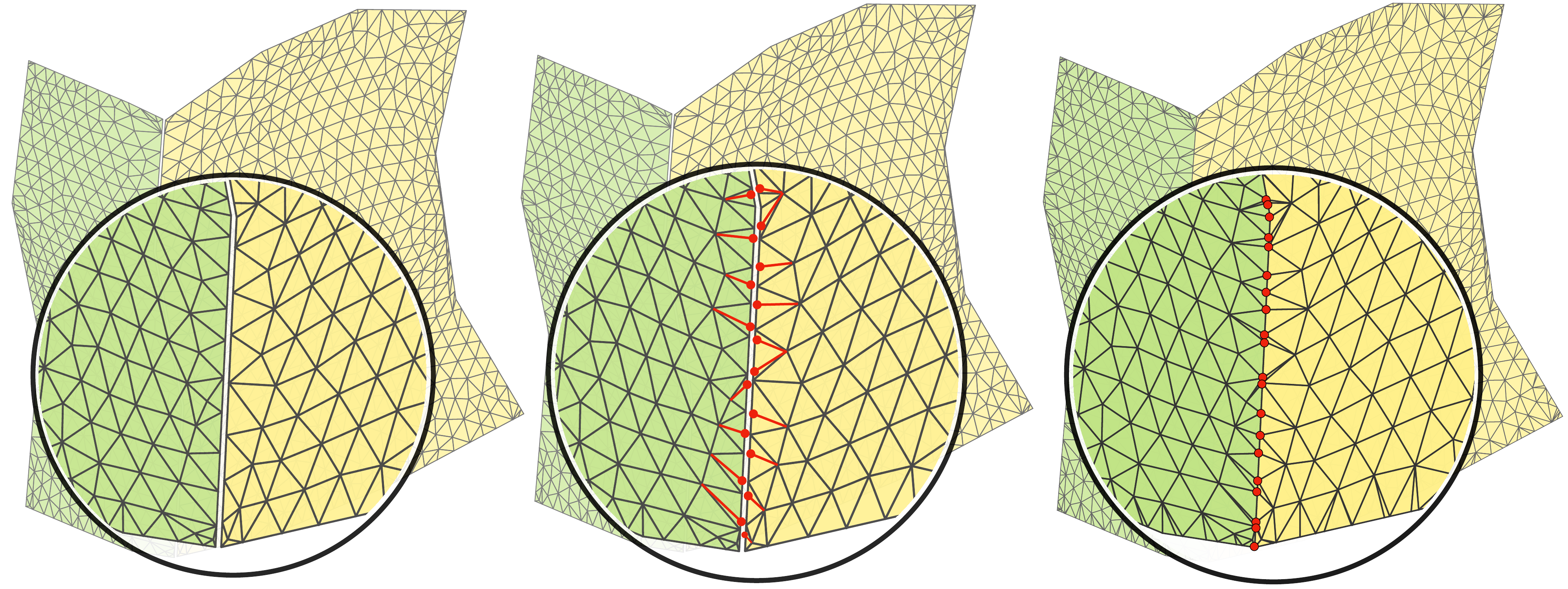}
     \caption{After embedding, a b-edge has a different number of vertices on the two sides. We refine it (red) to obtain a conforming mesh.}
    \label{fig:stich-ex1}
\end{figure}
Finally, to stitch the meshes together, we iterate over every b-edge~$e^b_j$, find the b-faces attached by it, and ``align'' the boundary edges of the two meshes. That is, we sort the union of the parametric values from both sides, and insert any missing vertex so that the two meshes are conforming on the boundary (\Cref{fig:stich-ex1}). This guarantees that adjacent meshes are conforming and share identical discretizations along each b-edge. As a result, the stitched mesh is globally conforming: each b-edge corresponds to a single shared polyline in the final mesh.
Note that when stitching edges from the same patch, the resulting mesh may become non-manifold if it is too coarse near the boundary. To prevent this issue, we refine the affected mesh edges twice, ensuring that there are always at least three edges between any pair of boundary edges.


\subsection{Stage 5. Remeshing}\label{sec:isotropic}

\begin{figure}
    \centering
    \includegraphics[width=0.48\linewidth]{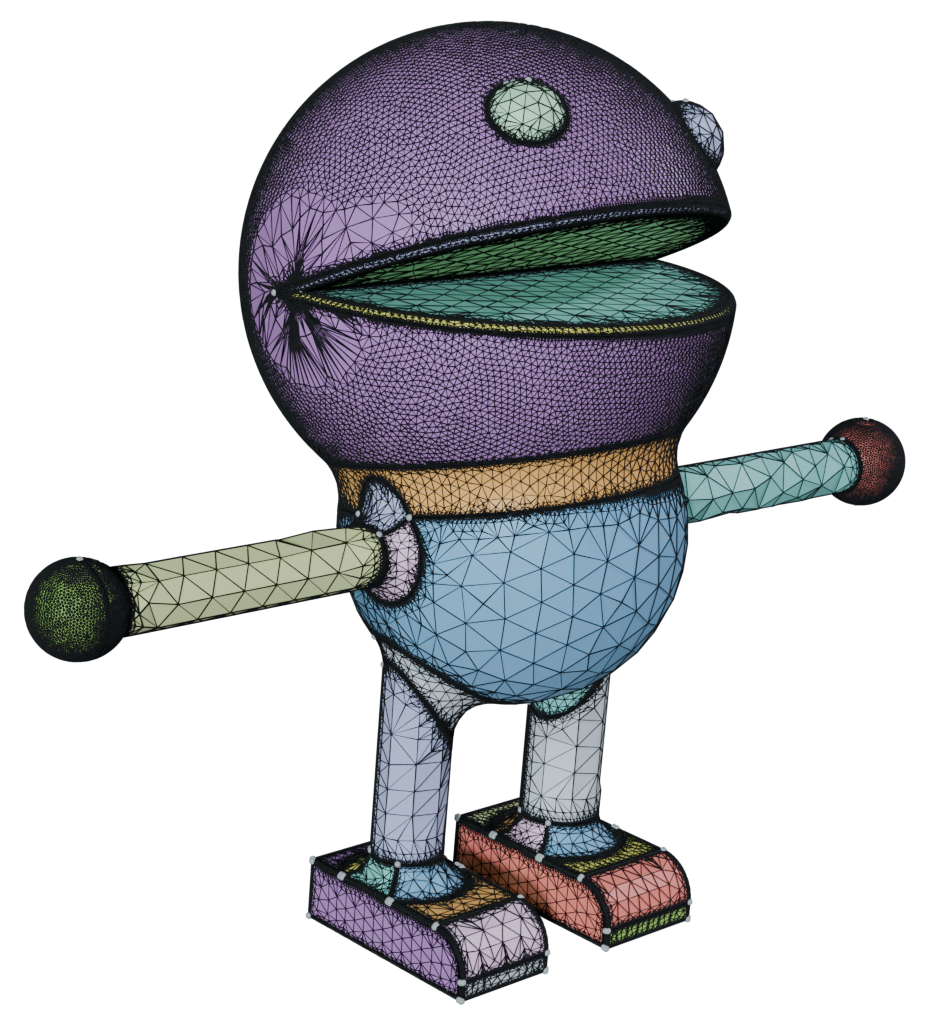}\hfill
    \includegraphics[width=0.48\linewidth]{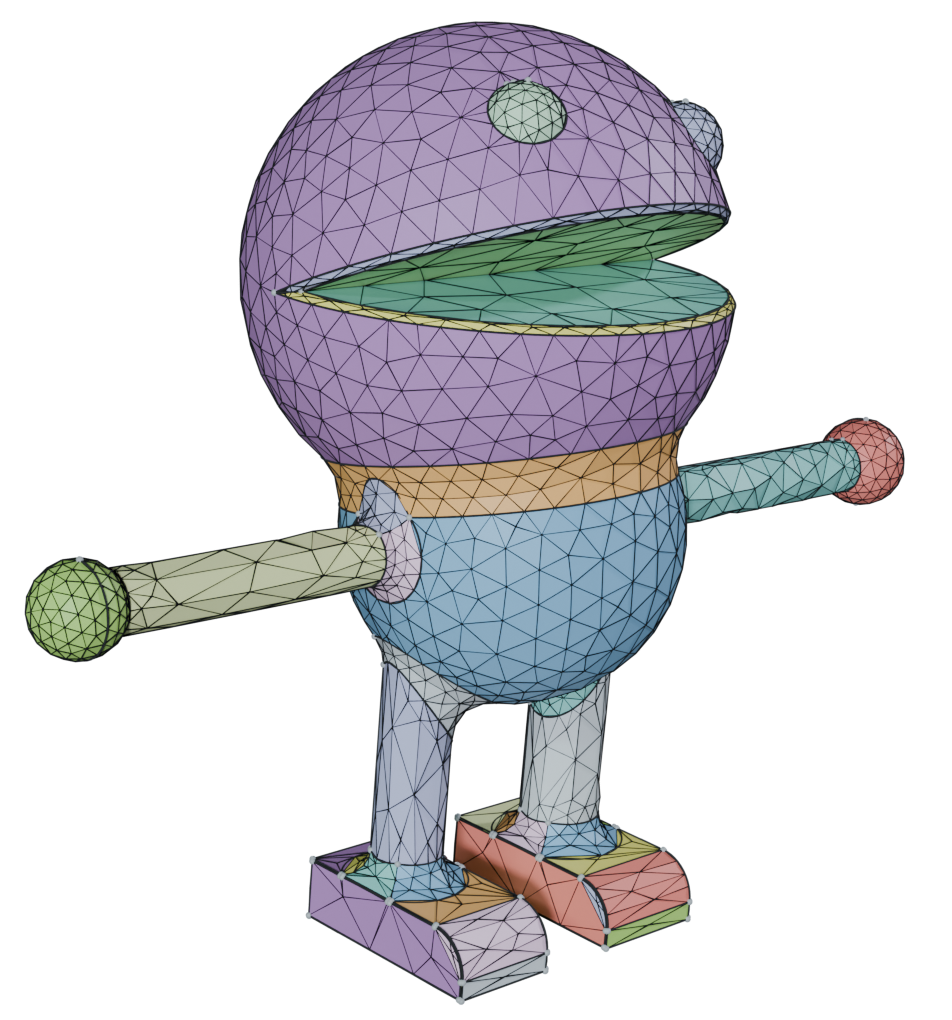}
    \caption{Example of a mesh before (left) and after remeshing (right).}
    \label{fig:meshing}
\end{figure}

The result of the previous stages is a mesh whose topology exactly matches the input B-Rep.
However, during projection and stitching, additional vertices may be inserted, leading to irregular element sizes and reduced mesh quality (Figure~\ref{fig:meshing}, left).
To improve mesh quality, we apply isotropic remeshing~\cite{botsch2004remeshing} with several targeted modifications.

\paragraph{Topology Preservation}
To preserve our topological guarantees, we restrict remeshing operations to prevent any changes in mesh connectivity across b-edges, b-loops, and b-faces.
Specifically, using the declarative specification framework~\cite{wildmeshing}, we freeze b-vertices, prevent edge swaps on b-edges, and allow edge collapses only on internal edges or along b-edges (but never between them).

\paragraph{Geometry Accuracy}
Inspired by TetWild~\cite{hu2018tetrahedral,hu2020fast}, we constrain vertex motion using two geometric envelopes: one enforcing proximity to the parametric surfaces associated with b-faces, and one enforcing proximity to the 3D curves associated with b-edges.
This ensures that the remeshed surface remains within a user-specified tolerance of the input geometry.

\paragraph{Surface Smoothing}
Isotropic remeshing typically relies on tangential smoothing to improve element quality while avoiding surface shrinkage.
In our setting, since vertices must remain on the input geometry, we first apply classical Laplacian smoothing and then project onto the corresponding parametric surface.
This effectively mimics tangential smoothing on the input surface while respecting the geometric envelopes.

\paragraph{Quality Optimization}
Isotropic remeshing aims to produce regular meshes, often favoring vertex valences close to 6 by performing edge flips.
In our case, rather than enforcing a target valence, we explicitly evaluate triangle shape regularity~\cite{bank1997mesh} before and after each edge flip and accept the operation only if it improves it.

\paragraph{Degenerate Triangles}
To prevent the creation of degenerated triangles, we disallow any operation that would introduce triangles with zero area.

\paragraph{Fold-overs}
We prevent fold-overs by rejecting an edge flip if it increases the dihedral angle between the adjacent to more than $\ang{120}$ by checking if the dot product of the two normals is less than~$-0.5$.

\subsection{Heuristic}\label{sec:acceleration}

The previously described algorithm produces a mesh with the same topology as the input B-Rep; however, the embedding procedure, the subsequent projections, and stitching may deviate significantly from the B-Rep geometry. To improve both runtime and geometric accuracy, we introduce several heuristic accelerations (\Cref{sec:ablation}). These heuristics do not affect correctness: if either heuristic fails its validation checks, we automatically revert to the topology-preserving correct algorithm. Note that all heuristic outputs are validated solely on topological criteria before acceptance.

\paragraph{Initial refinement} 
To ease the tracing of b-edges (Section~\ref{sec:tracing}), we apply an additional RGB subdivision.
For every b-edge $e^b_i$ attached to a b-face $f^b_j$, we upsample its associated curve using $0.01\epsilon$, with at least 10 sample points.
All upsampled points are then projected to their closest points on the mesh $M_j$, and we apply RGB subdivision until each triangle contains at most one projected sample point.

\paragraph{Singular patches}
If the patch has a singularity on the boundary of the parametric domain, we collapse all boundary edges to a single point. This allows us to create meshes with low geometric error for shapes such as spheres and cones.

\paragraph{Optimistic tracing}
During tracing of inner loops, we first try to embed them without connecting existing loops to the boundary, assuming that the embedded loop partitions the mesh into two connected components, one of which is topologically a disk. If this condition is satisfied, the trace is accepted and the algorithm proceeds. If the embedded cycle does not yield such a partition, we reject the trace and apply the standard algorithm.

\paragraph{Outer loop tracing}
In our experience, most outer loops coincide with the boundary of the parametric patch.
For this reason, embedding the outer loop while excluding boundary edges can introduce large geometric distortion and unnecessary refinement.
To limit this issue, we allow the outer loop to use boundary edges during tracing.
After each vertex is connected, we verify that the mesh remains a single connected component, as the trace may inadvertently connect two distinct boundary edges of the meshed patch.
If this condition is violated, we project all remaining points, discard the component with the fewest projected points, and continue tracing on the remaining component using the standard procedure.
Once the outer loop has been fully traced, we verify that the resulting mesh is topologically a disk; if this check fails, we revert and use the default outer-loop tracing algorithm.

\begin{figure}
    \centering
    \includegraphics[width=\linewidth]{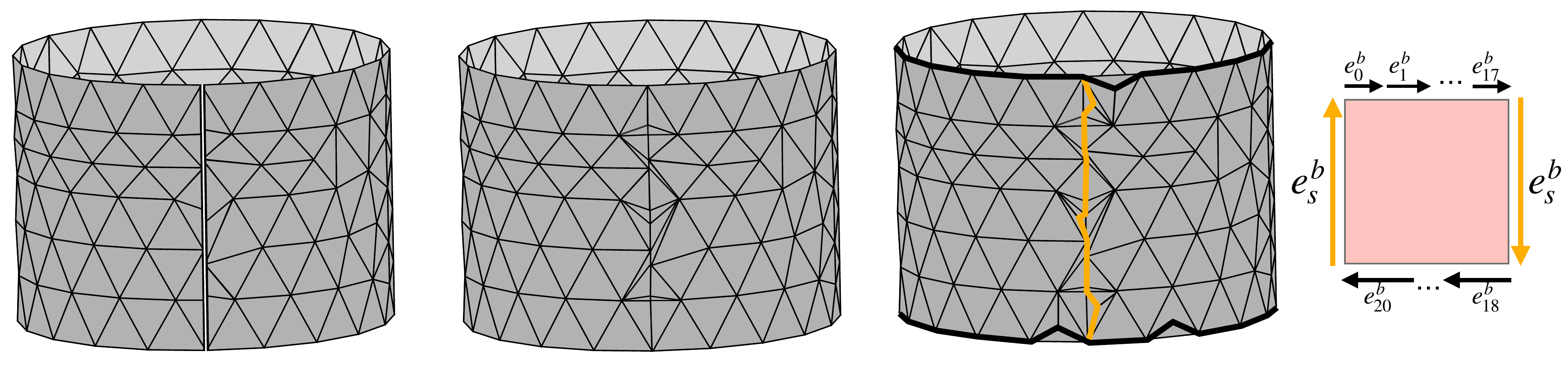}
    \caption{Handling periodic patches. We first stitch the mesh (double lines) to form a conforming periodic surface. We then trace b-edges as usual, while avoiding tracing repeated b-edges (orange) more than once.}
    \label{fig:periodic}
\end{figure}
\paragraph{Periodic patches}
Since we rely on closest-point projection, when two regions of $M_k$ are close in $\mathbb{R}^3$ but far apart along the surface or not adjacent in the mesh connectivity, as is common for periodic b-faces, the projection may alternate between geometrically nearby but topologically distant locations, producing a ``zig-zag'' (\Cref{fig:ablation}, bottom).
Additionally, repeatedly tracing identical b-edges can lead to unnecessary mesh refinement and poor geometric quality.
We therefore devise a heuristic that avoids retracing vertices and edges that have already been embedded.
If the parametric surface associated with the b-face is periodic, we first modify the topology of $M_k$ to match the surface periodicity (\Cref{fig:periodic}).
We then trace the outer loop as usual, except that repeated b-edges are traced only once.
After tracing, we require that the retained component is topologically an open disk.
If this condition is not satisfied, we revert the trace, restore the original disk topology of $M_k$, and trace the outer loop using the standard algorithm.
Note that this heuristic is applied only to outer loops, since after tracing the outer loop, the remaining component is a disk.

\paragraph{Long traces}
During loop tracing, if the length of the embedded edge-chain exceeds the length of the corresponding b-edge by more than a factor of 100, we stop the trace and resample the patch with a smaller tolerance. This heuristic is used purely for performance: when the mesh is too coarse, no sufficiently close path may exist on the surface.
Refining the mesh increases resolution and typically enables a shorter, more accurate trace.

%% file: 40-implementation.tex

%% file: 50-results.tex
\section{Results}
\label{sec:results}

Our algorithm is implemented in Python and C++, and uses the Wildmeshing-toolkit~\cite {wildmeshing} for mesh data structures and editing operations.
All experiments were run on a single cluster node equipped with an Intel Cascade Lake Platinum 8268 and Xeon Platinum 8592 processor, limited to one thread, 100\,GB of RAM, and a maximum runtime of 8 hours per model.


\subsection{Datasets and experimental setup}

We evaluate our method on real-world CAD data drawn from two datasets: one chunk of the Fusion360~\cite{willis2021joinable} dataset (approximately 750 models) and one chunk of the ABC~\cite{Koch_2019_CVPR} dataset (approximately 10\,000 models).
Since both datasets provide models in STEP format, we use the B-Rep representations from the ABS dataset~\cite{abs}, which contains a one-to-one conversion of STEP files into an HDF5 format suitable for processing.
When a model contains multiple parts, we mesh only the first part.
These datasets contain complex B-Reps with a wide range of geometric and topological configurations, including thin features, small tolerances, and inconsistent geometric embeddings. Importantly, the data originates from different CAD kernels: Fusion360 uses its own kernel, while ABC models are sourced from OnShape, which uses the Parasolid kernel.

Only 317 (i.e., around 3\%) models did not terminate within the imposed time and memory limits.
In all such cases, increasing the available resources allowed the computation to complete successfully, indicating that failures are due to resource limits rather than algorithmic breakdowns.

\begin{figure}
    \centering
    \includegraphics[width=\linewidth]{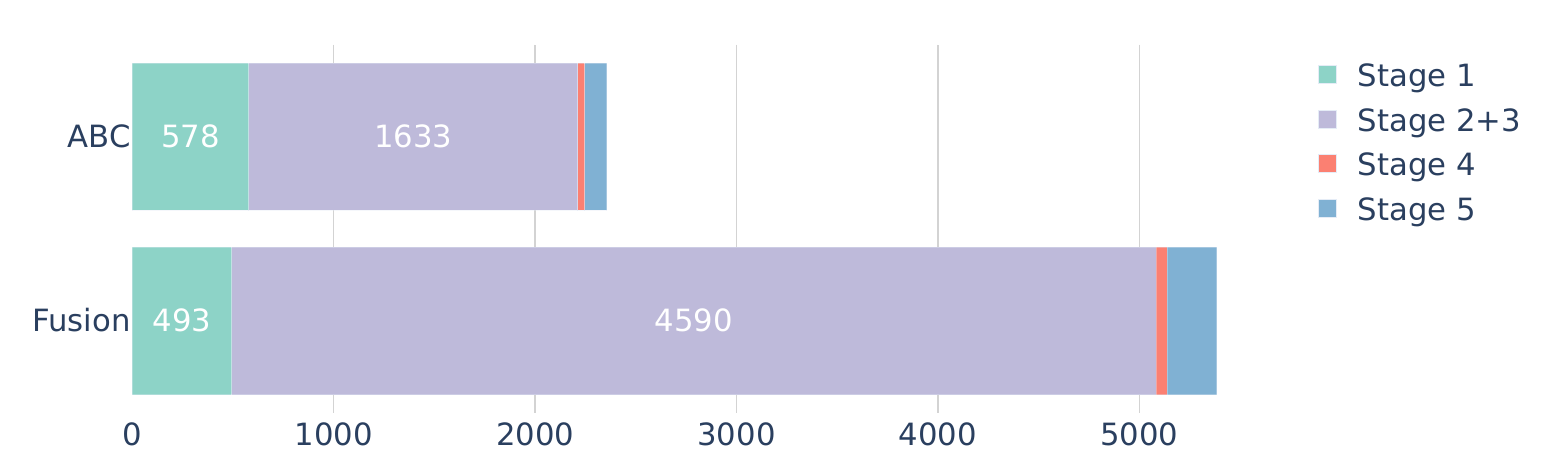}
    \caption{Average runtime distribution of the individual stages for the two datasets.}
    \label{fig:time}
\end{figure}

\paragraph{Runtime}
Figure~\ref{fig:time} reports the runtimes of the individual stages for the two datasets. Stages 2 and 3 were grouped, as the runtime of stage 2 is negligible.
Fusion360 models generally require longer runtimes due to their greater geometric and topological complexity.
A breakdown of the runtime shows that loop embedding dominates the overall cost (70\% and 85\% of the total runtime for ABC and Fusion, respectively), followed by sampling (25\% and 10\%), while stitching and remeshing contribute marginally.


\begin{figure}
    \centering
    \footnotesize
    \includegraphics[width=0.49\linewidth]{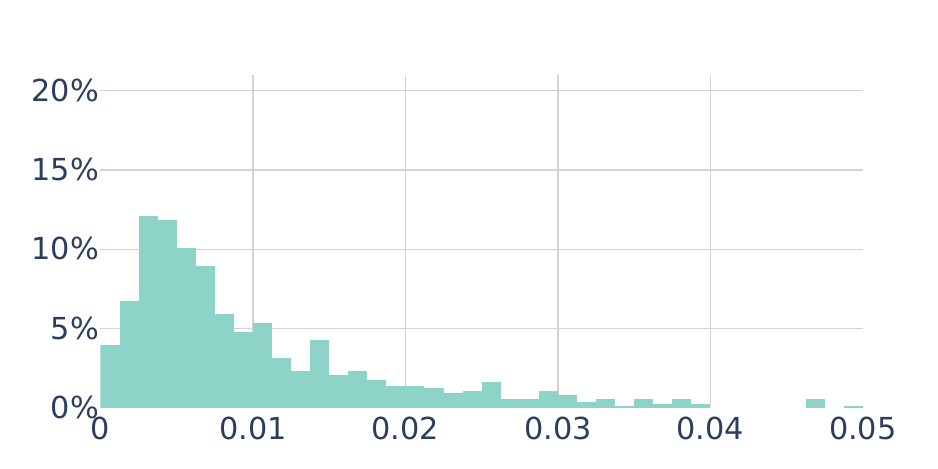}\hfill%
    \includegraphics[width=0.49\linewidth]{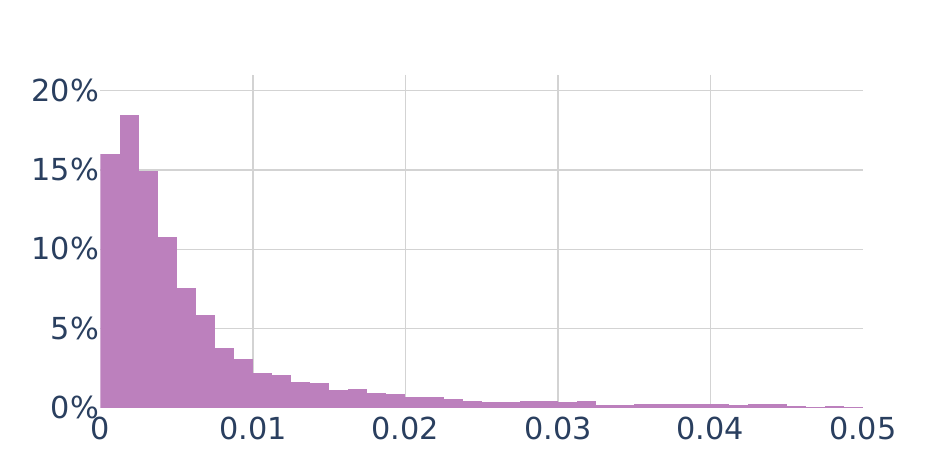}\\
    \parbox{0.49\linewidth}{\centering Fusion}\hfill%
    \parbox{0.49\linewidth}{\centering ABC}
    \caption{Histogram of the geometric error relative to the bounding box diagonal.}
    \label{fig:geom-error}
\end{figure}
\paragraph{Geometric accuracy}
Although our method does not guarantee geometric optimality (snapping operations introduce local distortion), it consistently produces meshes that closely approximate the input geometry.
Figure~\ref{fig:geom-error} shows a histogram of the geometric error normalized by the bounding box diagonal.
More than 93\% of the B-Rep models exhibit errors below 3\%, whereas only 1\% exhibit deviations larger than 10\% up to 38\%.

\subsection{Ablation study}\label{sec:ablation}

\begin{figure}
    \centering\footnotesize
    \includegraphics[width=\linewidth]{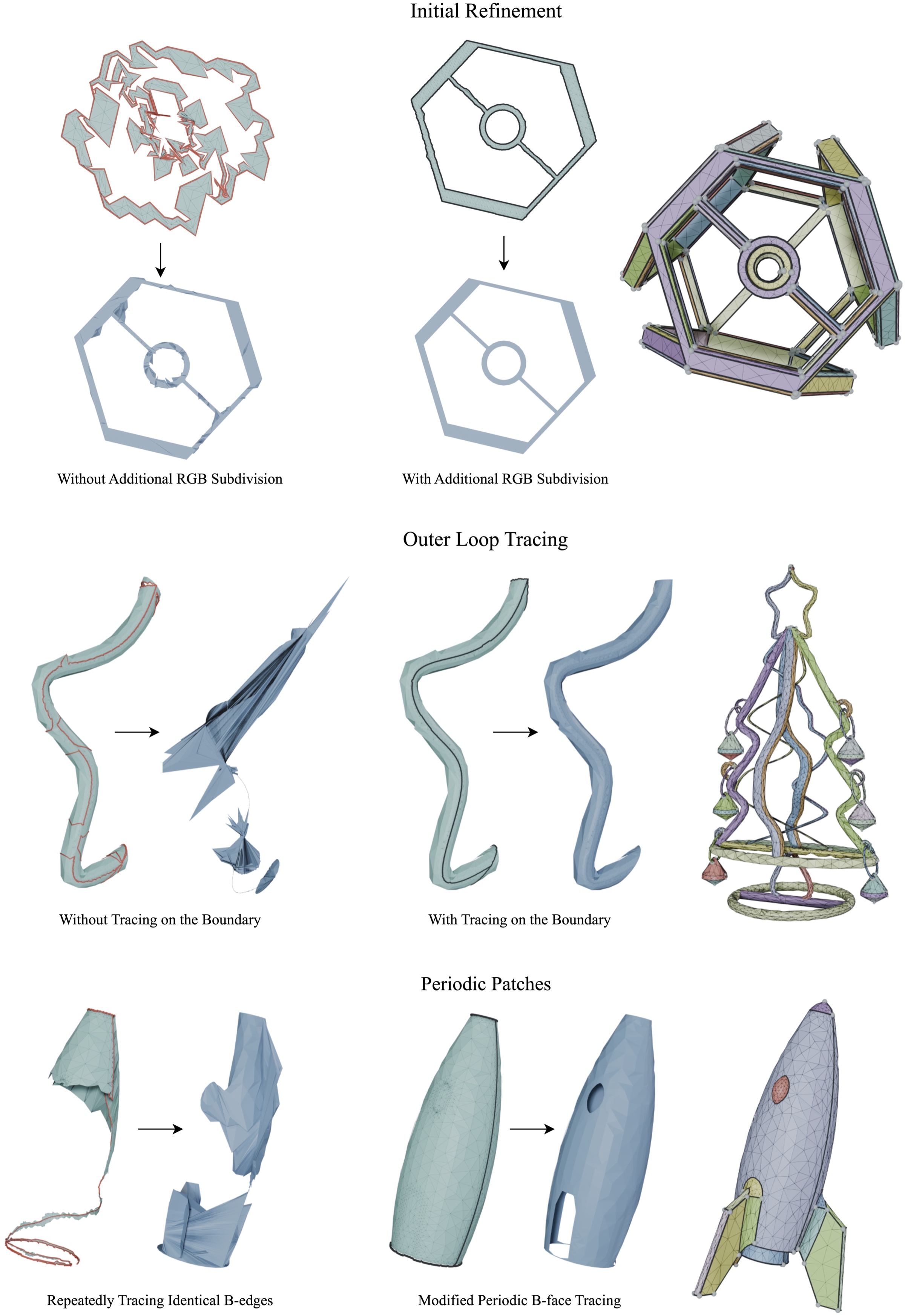}
    \caption{Example of a B-Rep model run with and without heuristics. Both meshes are topologically valid; however, using the full pipeline yields a more geometrically accurate model.}
    \label{fig:ablation}
\end{figure}

To evaluate the impact of our heuristic accelerations, we perform an ablation study by disabling them.
We compare the runtime and results with and without the heuristic.
Disabling heuristics significantly lowers geometric accuracy, particularly for models with periodic faces or closely spaced surface regions, while having little effect on topological correctness. With the whole pipeline, our method successfully generates a high-quality mesh (Figure~\ref{fig:ablation}).


\begin{figure*}
    \centering
    \includegraphics[width=\linewidth]{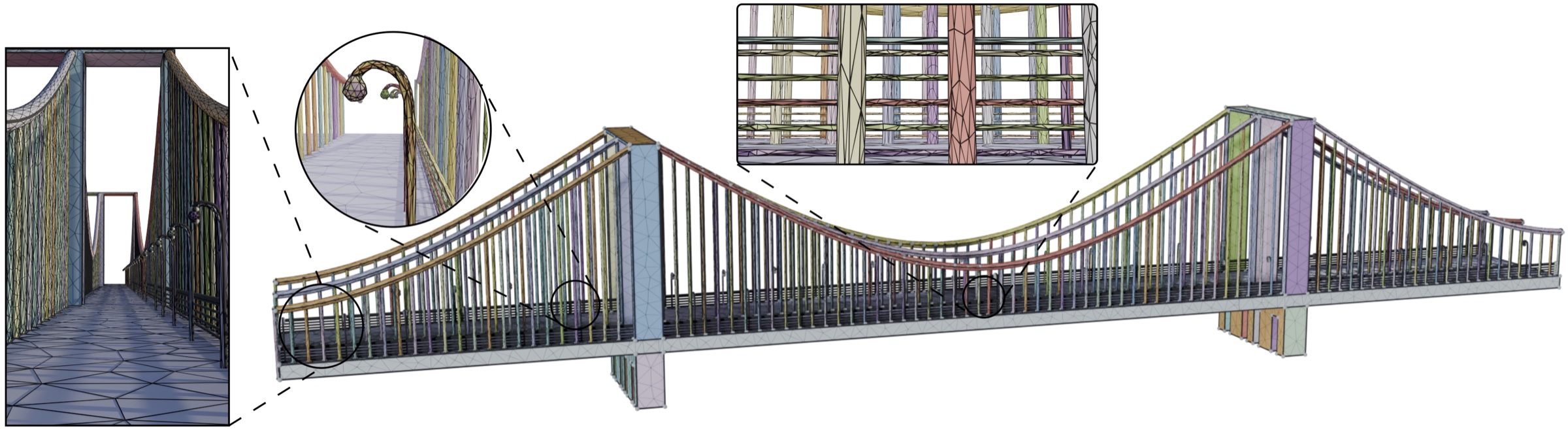}
    \caption{A large B-Rep model with fine details is correctly meshed by our method.}
    \label{fig:bridge}
\end{figure*}

\subsection{Robustness}

\paragraph{Scale variation}
Our method successfully handles models with extreme scale variation, thin features, and complex loop structures.
Figure~\ref{fig:bridge} shows a large-scale engineering model containing fine details spanning several orders of magnitude, which is correctly meshed by our method without manual intervention.

\begin{figure}
    \centering
    \includegraphics[width=\linewidth]{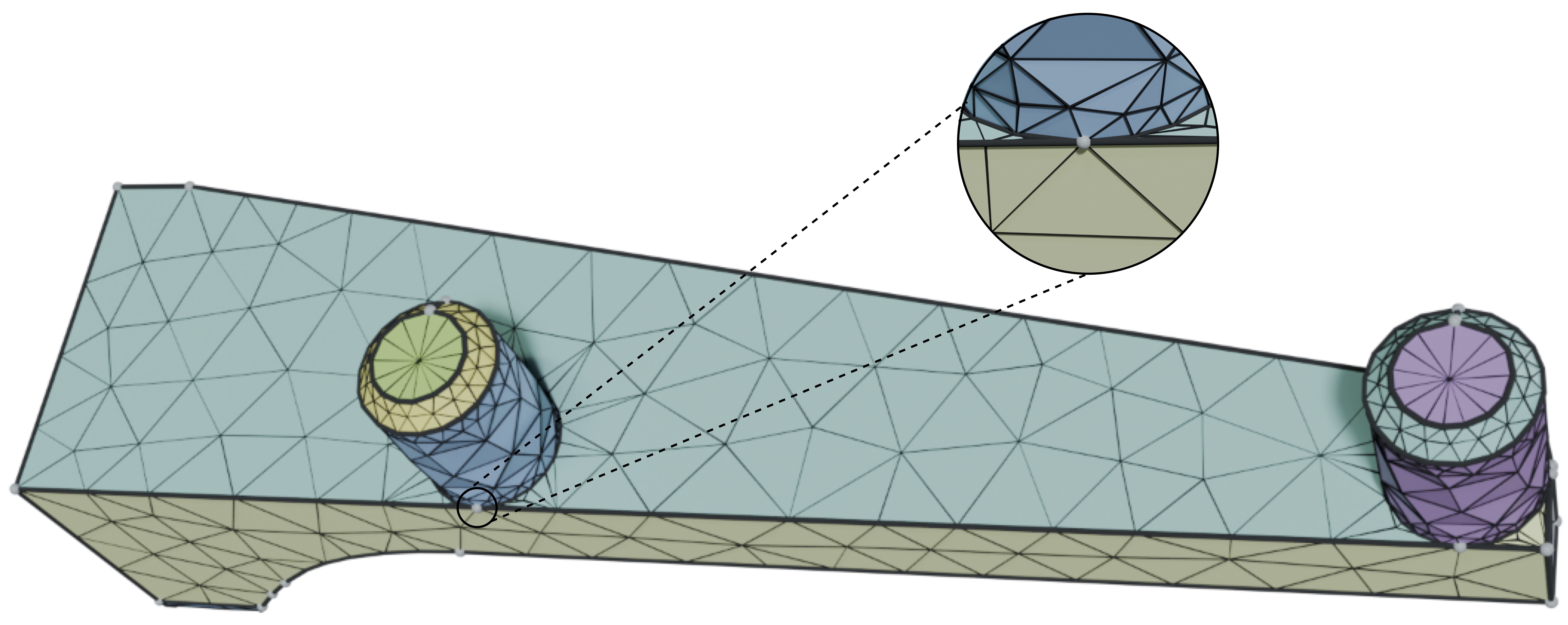}
    \caption{B-Rep with non-manifold b-loops successfully meshed by our method.}
    \label{fig:nonmanifold}
\end{figure}

\paragraph{Topology}
Our method robustly handles non-manifold loop configurations commonly found in real B-Rep data (Section~\ref{sec:tracing}).
Figure~\ref{fig:nonmanifold} illustrates an example in which a single b-vertex participates in multiple loops on the same patch.
Our method enables processing of any model, regardless of b-loop topology, without relying on geometric repair heuristics.

\begin{figure}
    \centering
    \includegraphics[width=\linewidth]{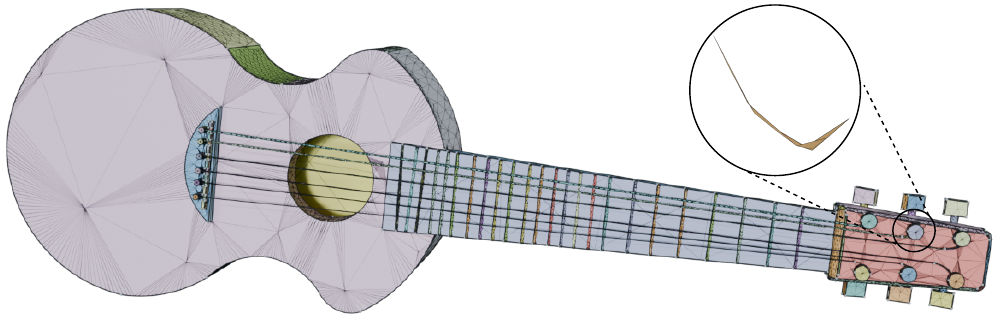}
    \caption{B-Rep with extremely small parametric domains (close-up) successfully meshed by our method. Although several patches have nearly degenerate parameterizations, the resulting surface mesh remains valid and topologically correct.}
    \label{fig:small-param}
\end{figure}

\paragraph{Geometry}
In the evaluated datasets, we encounter models whose parametric domains have extremely small area (below $10^{-14}$) or are degenerate.
Such cases are particularly challenging for methods that rely on geometric tolerances or robust inversion of the parameterization.
Our method remains robust for these cases.
To avoid numerical issues during initial surface sampling, we pad the parametric domain by a small amount ($10^{-10}$) before meshing.
This padding enlarges the domain just enough to enable the construction of an initial triangle mesh without altering the intended topology.
All subsequent stages of the algorithm operate unchanged, and the padding does not affect the final mesh connectivity or correctness.
Figure~\ref{fig:small-param} shows an example of a model with near-degenerate parametric domains that is successfully meshed by our method.

\subsection{Comparison with existing methods}
\label{sec:comparison}

To the best of our knowledge, OpenCascade~\cite{occt}, Gmsh~\cite{Geuzaine2009}, NetGen~\cite{Schoeberl1997Netgen}, and Mefisto~\cite{mefisto} are the only open-source systems capable of directly meshing B-Reps in STEP format.
Note that OpenCascade fails to generate meshes for 1.56\% of models in the ABC dataset and 8.82\% of models in the Fusion dataset~\cite{abs}.
All methods require user-defined parameters; for each, we evaluate three configurations: coarse, default, and fine. 

Let $\ell$ denote the diagonal length of the model's bounding box.
For all methods, we define three parameter regimes (coarse, default, and fine) by decreasing or increasing a parameter by a factor of 5 relative to the default.
For Gmsh, we modify the min/max element size:
the default setting is $[0, 10^{22}]$,
the coarse setting is $[0.25\ell, 10^{22}]$,
and the fine setting is $[0, 0.01\ell]$.
For OpenCascade, we vary the linear deflection:
the default is $10^{-3}\ell$,
the coarse setting is $5 \cdot 10^{-3}\ell$,
and the fine setting is $2 \cdot 10^{-4}\ell$.
NetGen (via FreeCAD) provides predefined qualitative resolution settings;
we use \emph{Moderate} as the default, \emph{Very Coarse} for coarse, and \emph{Very Fine} for fine.
Mefisto (via FreeCAD) automatically estimates an optimal edge length $l$, which we use as the default setting.
For the coarse setting we use $5l$, while for the fine setting we use $l/5$.
For our method, we vary the target edge length:
the default is $0.05\ell$,
the coarse setting is $0.25\ell$,
and the fine setting is $0.01\ell$.

\input{figs/comparison/comparison}

Existing methods frequently fail on challenging models (Figure~\ref{fig:comparison}).
OpenCascade is sensitive to parameter choices and often misses features (e.g., the funnel of the coffee grinder model).
Gmsh fails to capture thin structures at default resolution and runs out of memory on complex models (e.g., the lamp).
Mefisto performs better overall but produces incomplete meshes at low and default resolutions.
NetGen failed to generate valid meshes for most of the tested models.
In contrast, our method succeeds on all tested models, consistently preserving all B-Rep faces, loops, and adjacencies.

We also tested all methods on a simpler, smoother model (the seahorse), for which all methods generated a mesh.
At coarse resolution, Gmsh fails to capture curve features, while the other methods generate low-quality meshes.
With our method, features are consistently captured at all resolutions, and the resulting meshes exhibit higher quality.

\begin{figure}
    \centering
    \includegraphics[width=.45\linewidth]{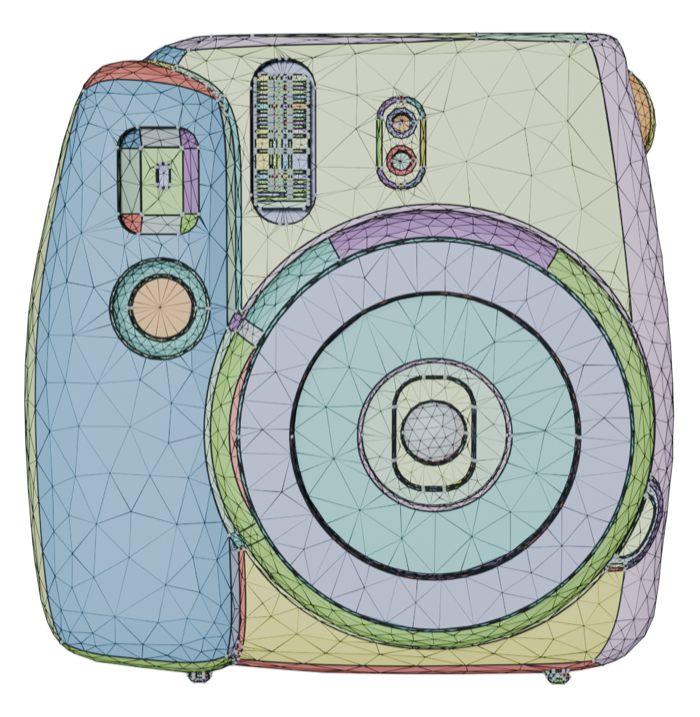}\hfill
    \includegraphics[width=.45\linewidth]{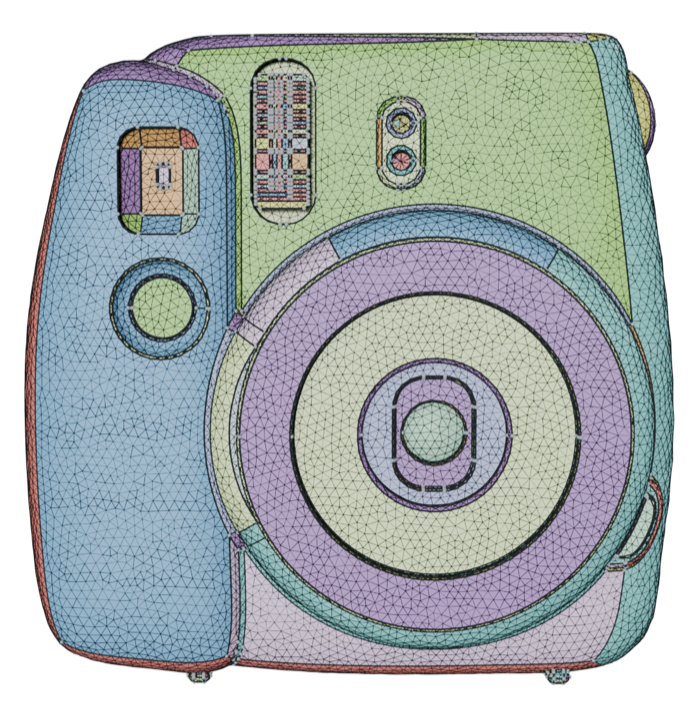}
    \includegraphics[width=.45\linewidth]{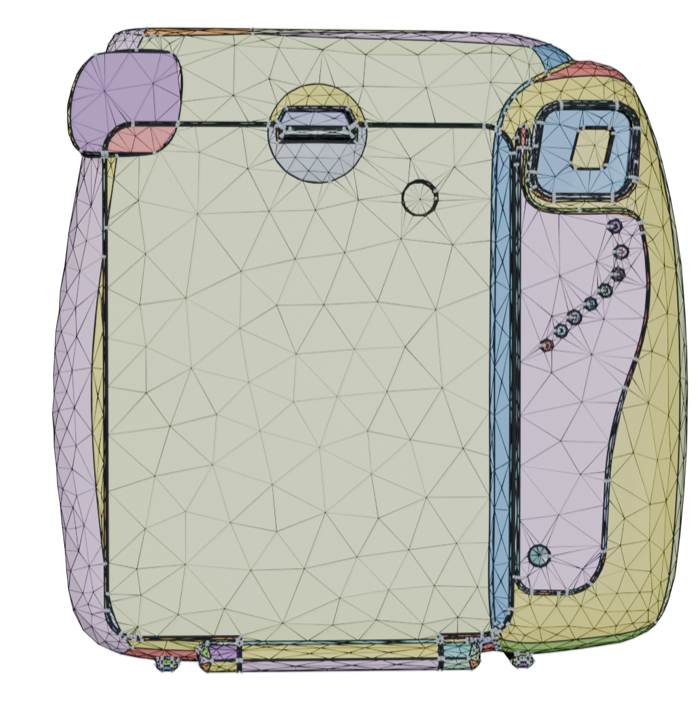}\hfill
    \includegraphics[width=.45\linewidth]{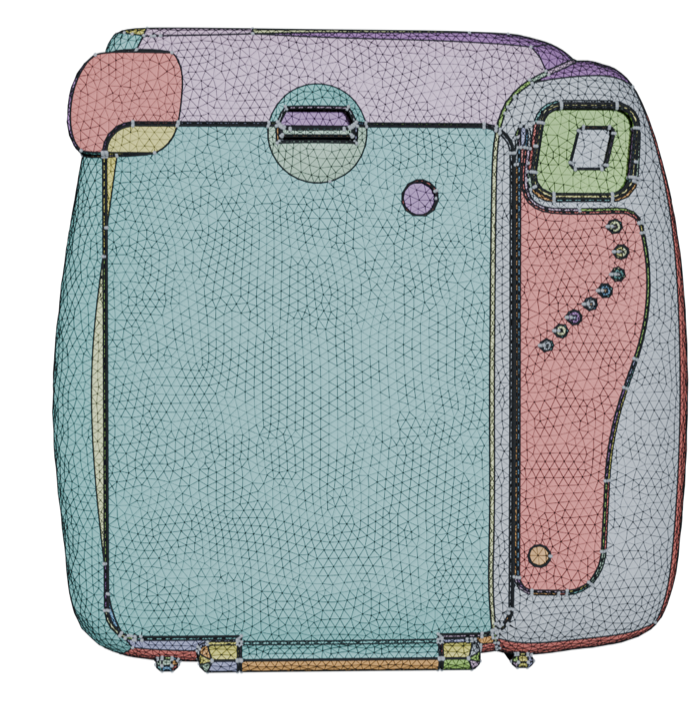}
    \caption{Examples of application for our coarse meshes. A mesh (left) can be refined (right) to increase its quality.}
    \label{fig:meshapp}
\end{figure}

\subsection{Downstream applications}
\begin{figure}
    \centering\footnotesize
    \includegraphics[width=\linewidth]{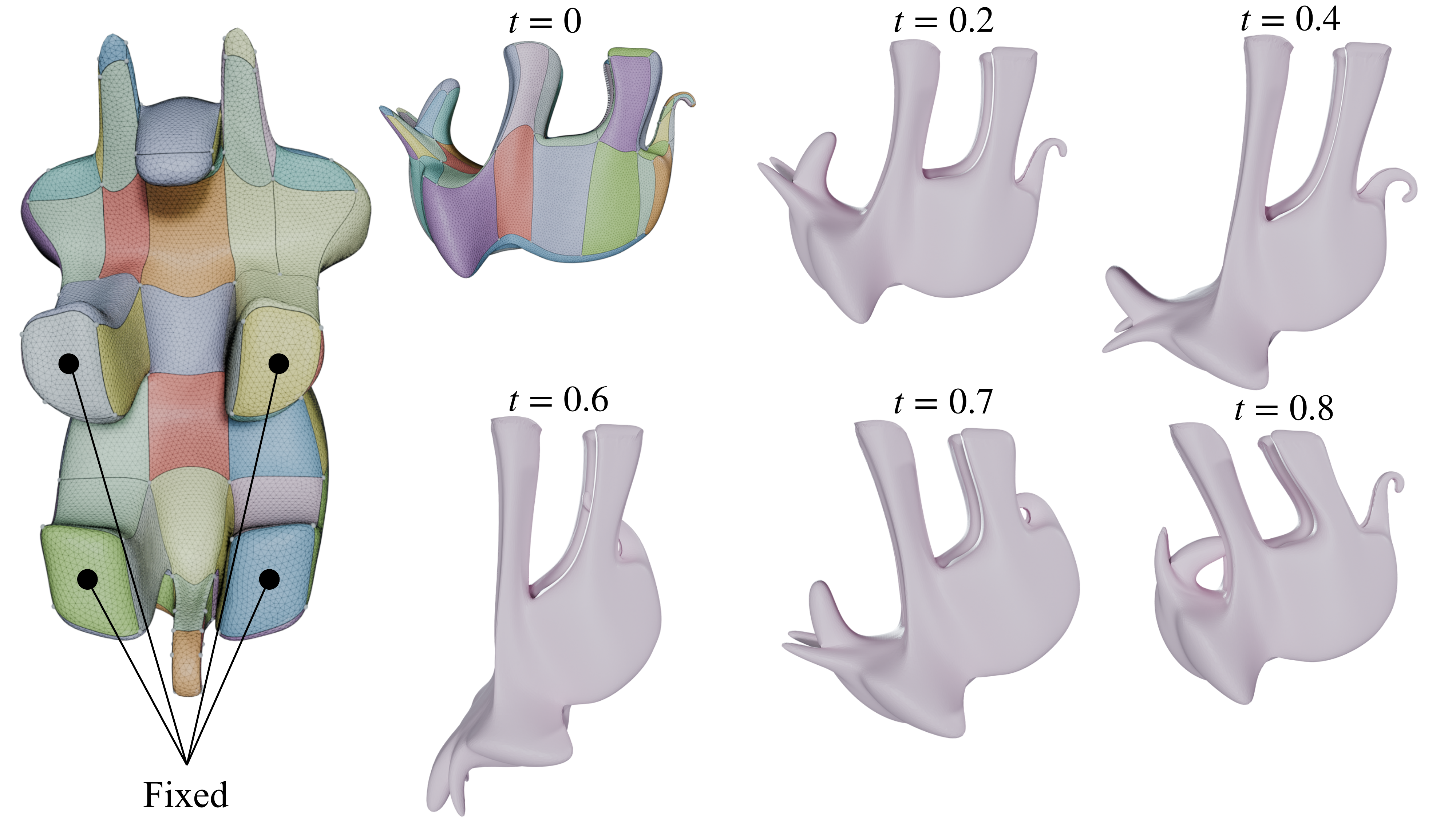}
    \caption{An input B-Rep is transformed into a simulation-ready tetrahedral mesh (left) and used directly in a finite element simulation with boundary conditions defined on B-Rep faces (right).}
    \label{fig:sim}
\end{figure}

The resulting meshes can be refined directly using standard surface-refinement techniques (Figure~\ref{fig:meshapp}). 
Additionally, our models can be used directly within a simulation pipeline in which boundary conditions are specified on B-Rep faces and automatically transferred to a tetrahedral mesh. We generate a volumetric mesh using TetGen~\cite{Si_2015} and propagate B-Rep face identifiers from our surface mesh to the tetrahedral boundary, enabling direct application of boundary conditions without any manual cleanup or retagging (Figure~\ref{fig:sim}).

%% file: figs/comparison/comparison.tex
\begin{figure*}
    \centering
\parbox{.01\linewidth}{~}\hfill
\parbox{.18\linewidth}{\centering gmsh}\hfill
\parbox{.18\linewidth}{\centering Mefisto}\hfill
\parbox{.18\linewidth}{\centering OCC}\hfill
\parbox{.18\linewidth}{\centering NetGen}\hfill
\parbox{.18\linewidth}{\centering Ours}


\parbox{.01\linewidth}{\rotatebox{90}{\centering Coarse}}\hfill
\parbox{.18\linewidth}{\includegraphics[width=\linewidth]{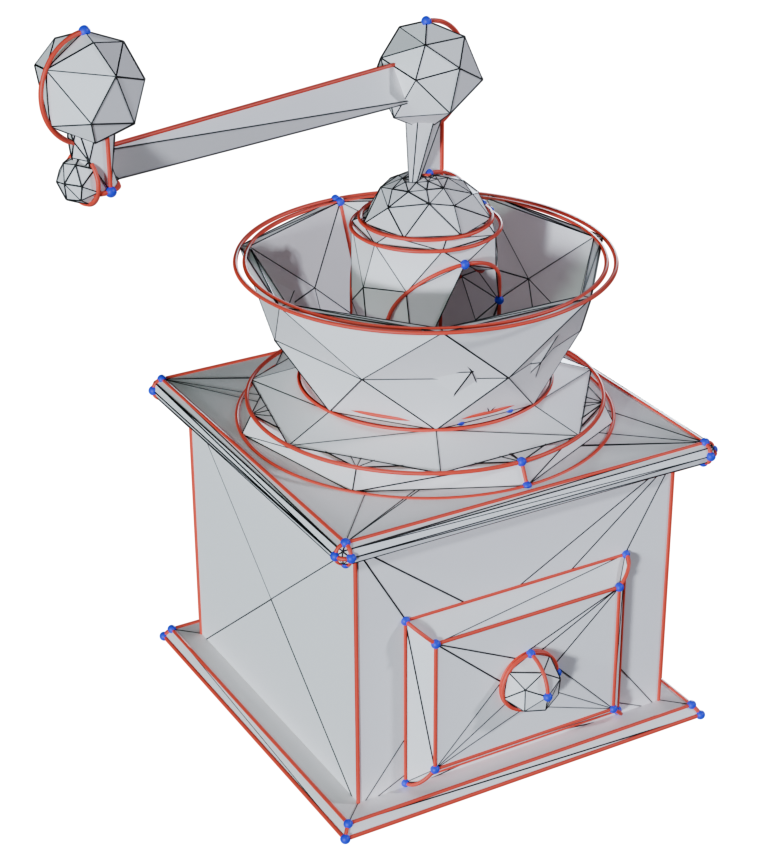}}\hfill
\parbox{.18\linewidth}{\includegraphics[width=\linewidth]{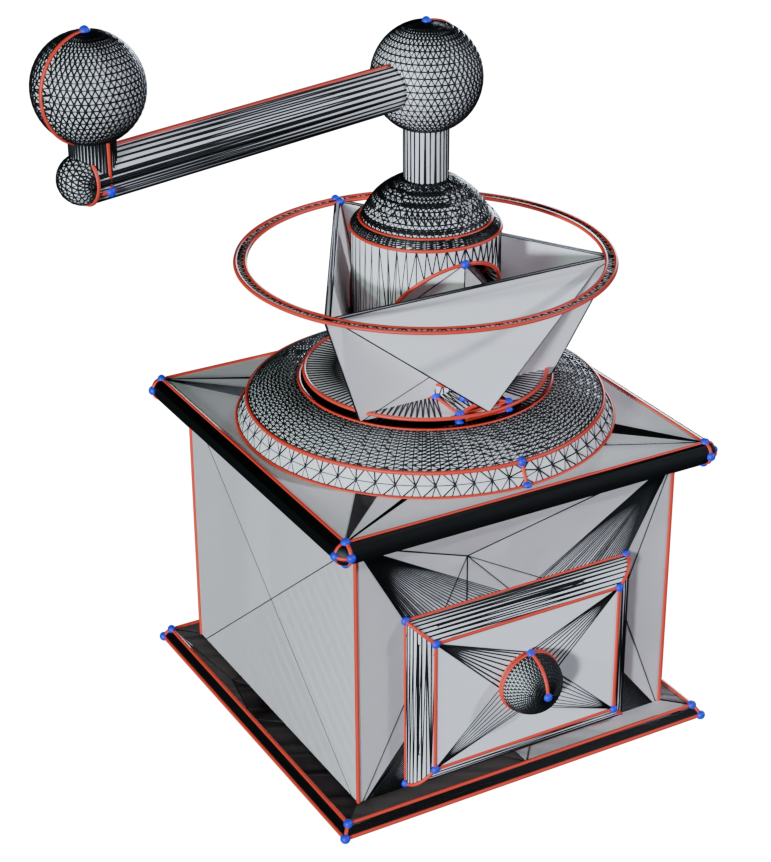}}\hfill
\parbox{.18\linewidth}{\includegraphics[width=\linewidth]{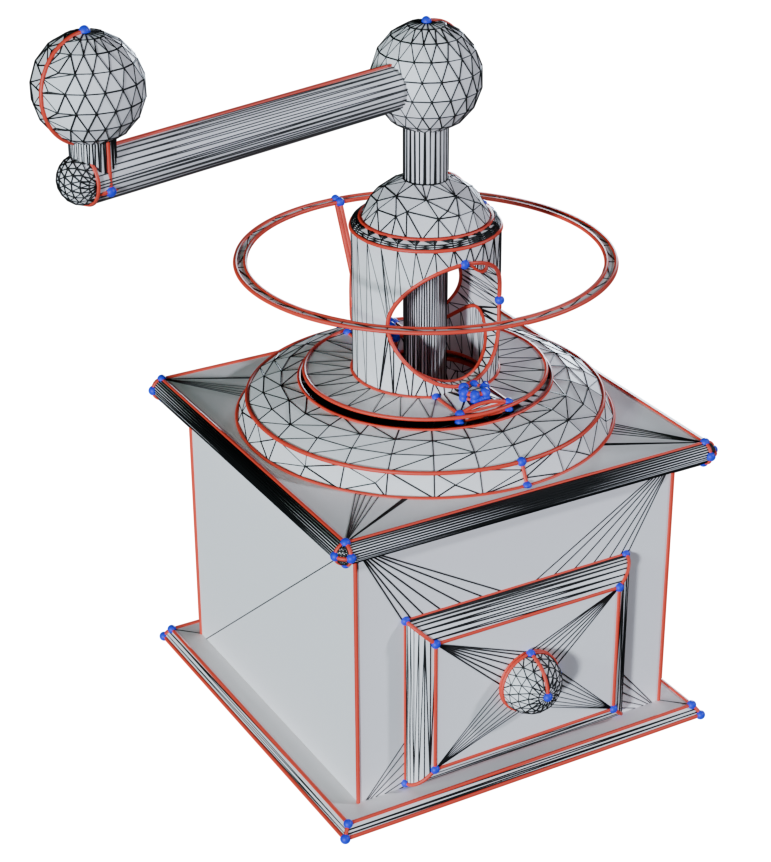}}\hfill
\parbox{.18\linewidth}{\centering Crash}\hfill
\parbox{.18\linewidth}{\includegraphics[width=\linewidth]{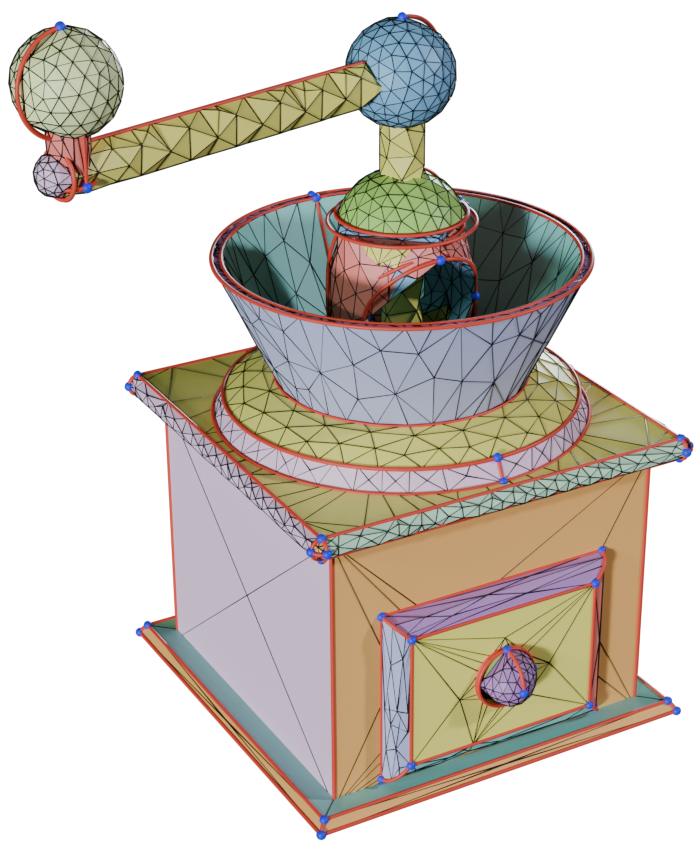}}

\parbox{.01\linewidth}{\rotatebox{90}{\centering Default}}\hfill
\parbox{.18\linewidth}{\includegraphics[width=\linewidth]{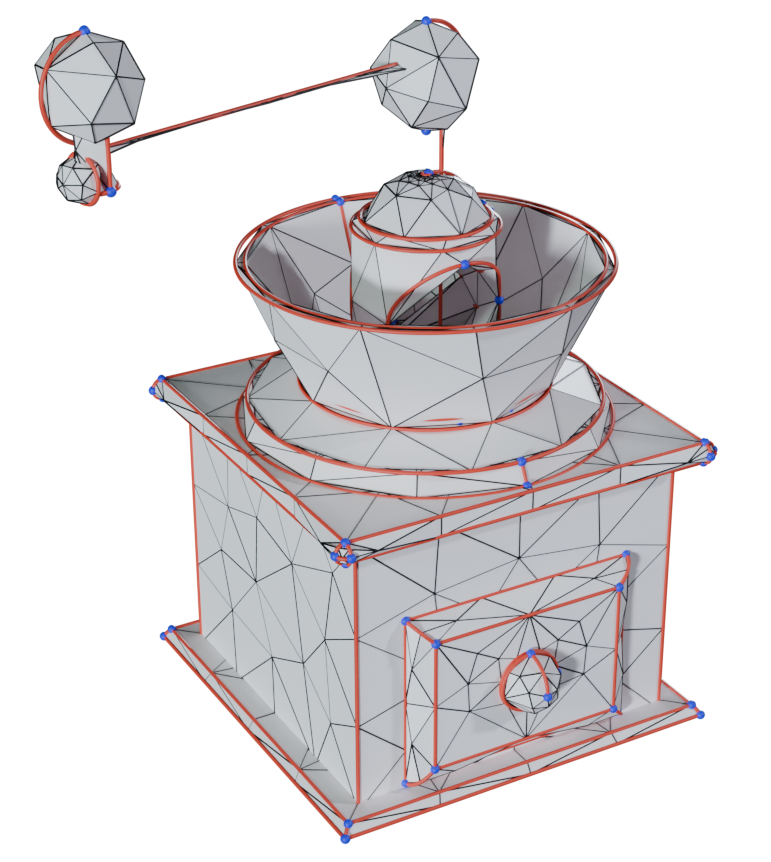}}\hfill
\parbox{.18\linewidth}{\includegraphics[width=\linewidth]{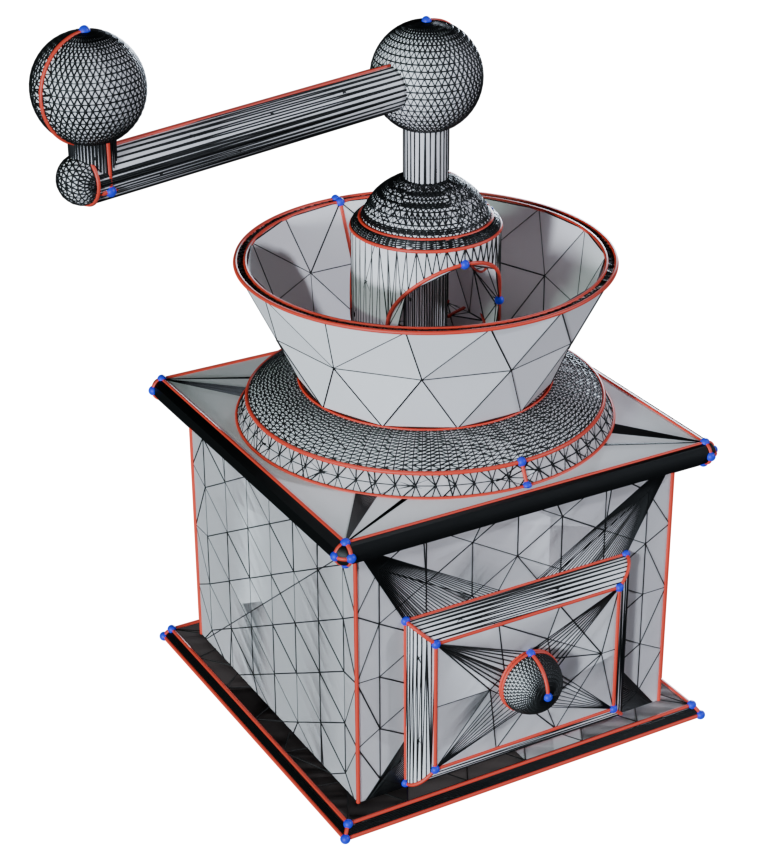}}\hfill
\parbox{.18\linewidth}{\includegraphics[width=\linewidth]{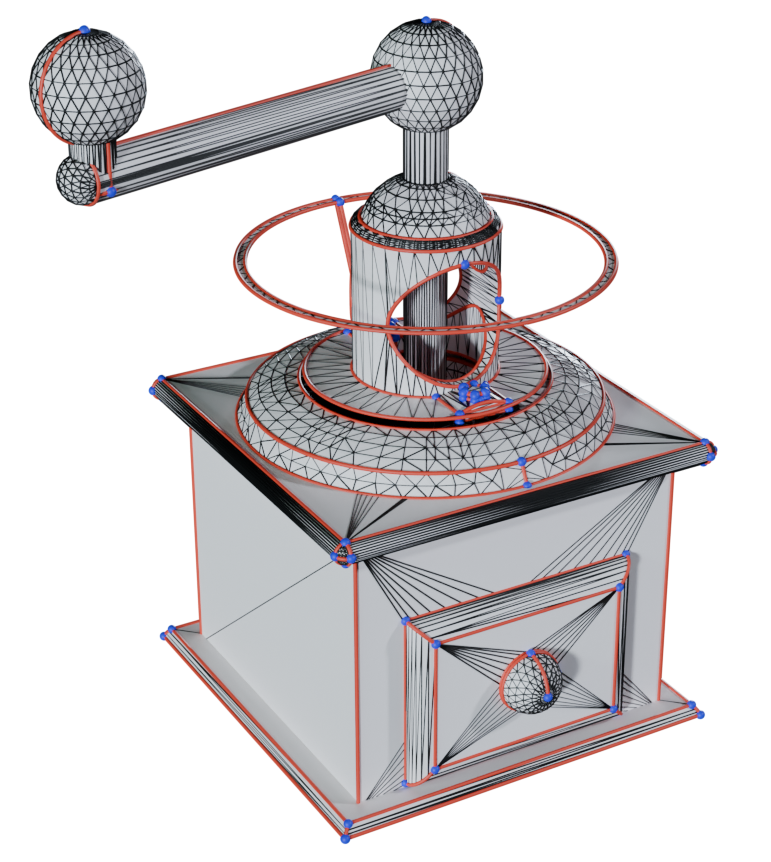}}\hfill
\parbox{.18\linewidth}{\centering Crash}\hfill
\parbox{.18\linewidth}{\includegraphics[width=\linewidth]{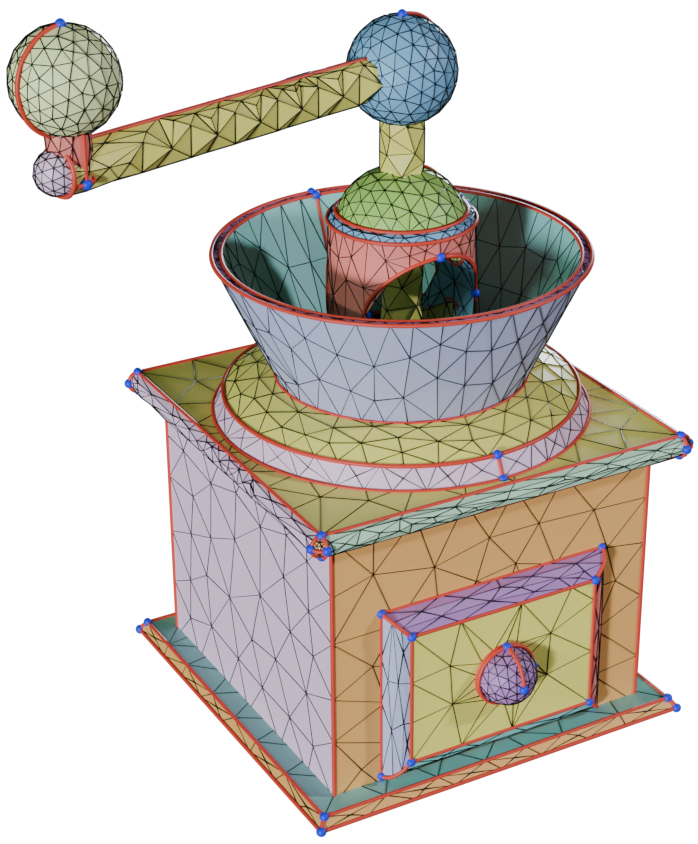}}

\parbox{.01\linewidth}{\rotatebox{90}{\centering Fine}}\hfill
\parbox{.18\linewidth}{\includegraphics[width=\linewidth]{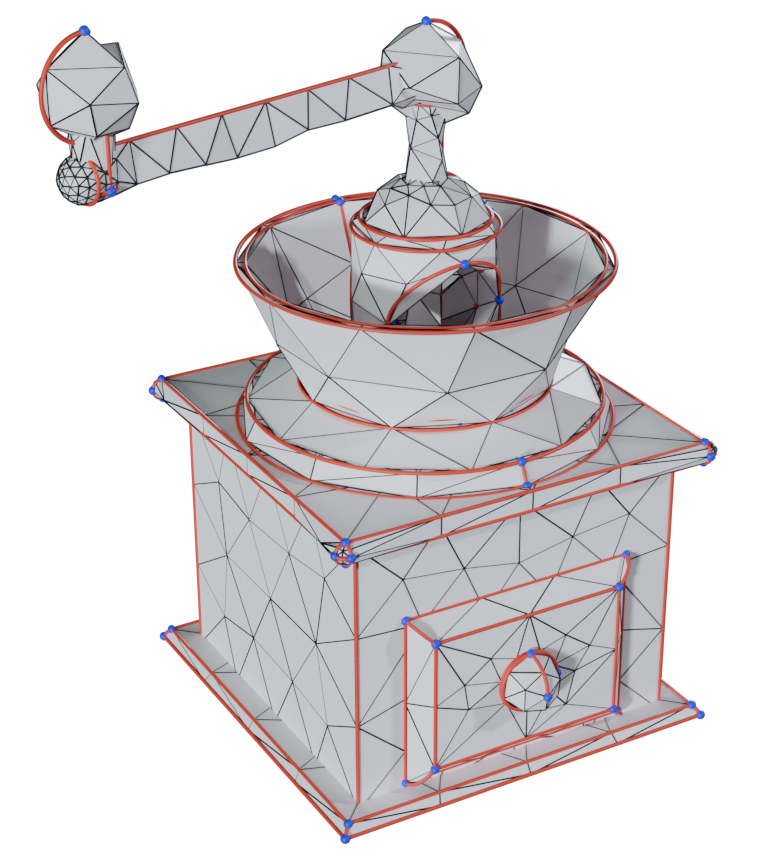}}\hfill
\parbox{.18\linewidth}{\includegraphics[width=\linewidth]{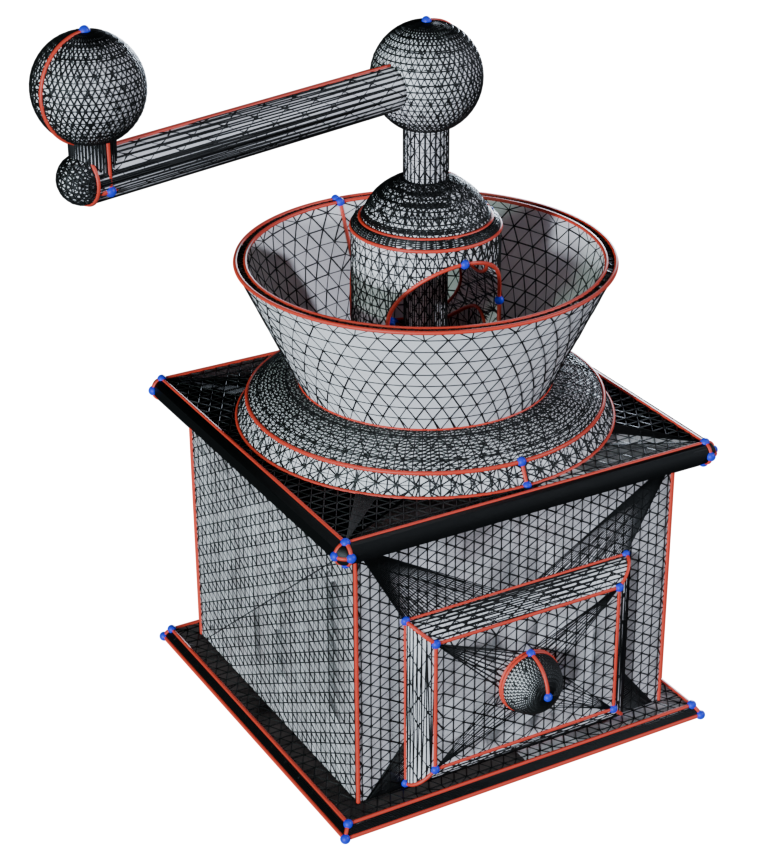}}\hfill
\parbox{.18\linewidth}{\includegraphics[width=\linewidth]{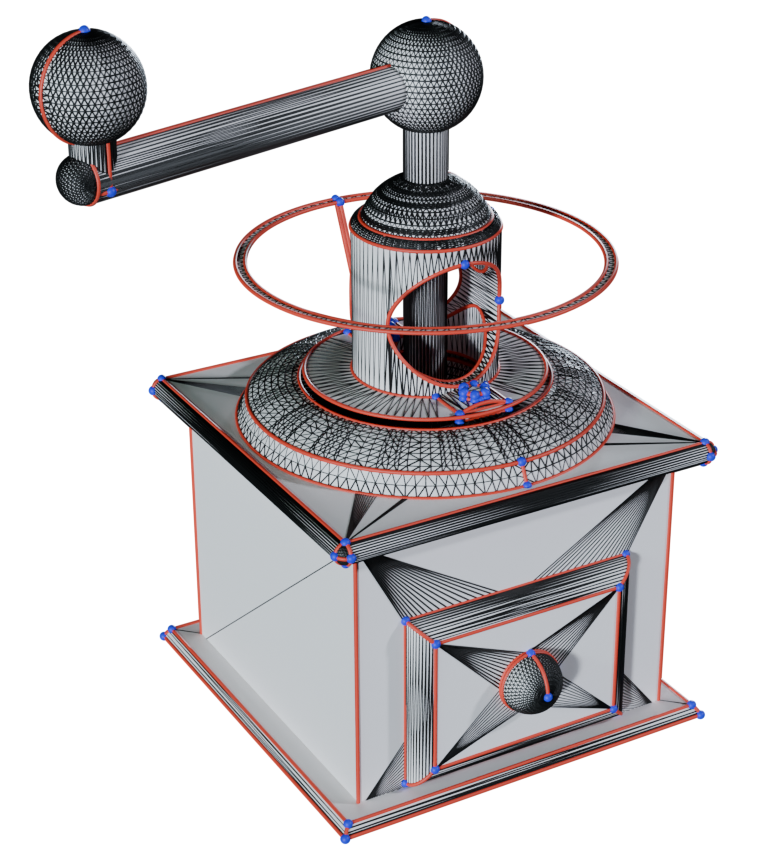}}\hfill
\parbox{.18\linewidth}{\centering Crash}\hfill
\parbox{.18\linewidth}{\includegraphics[width=\linewidth]{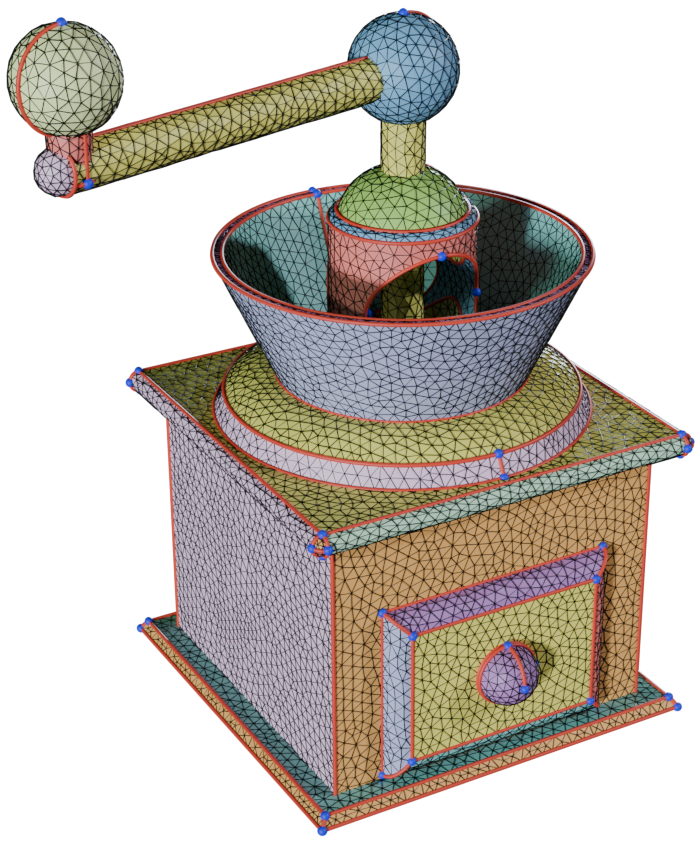}}

\parbox{.01\linewidth}{\rotatebox{90}{\centering Coarse}}\hfill
\parbox{.18\linewidth}{\centering Out of Memory}\hfill
\parbox{.18\linewidth}{\includegraphics[width=\linewidth]{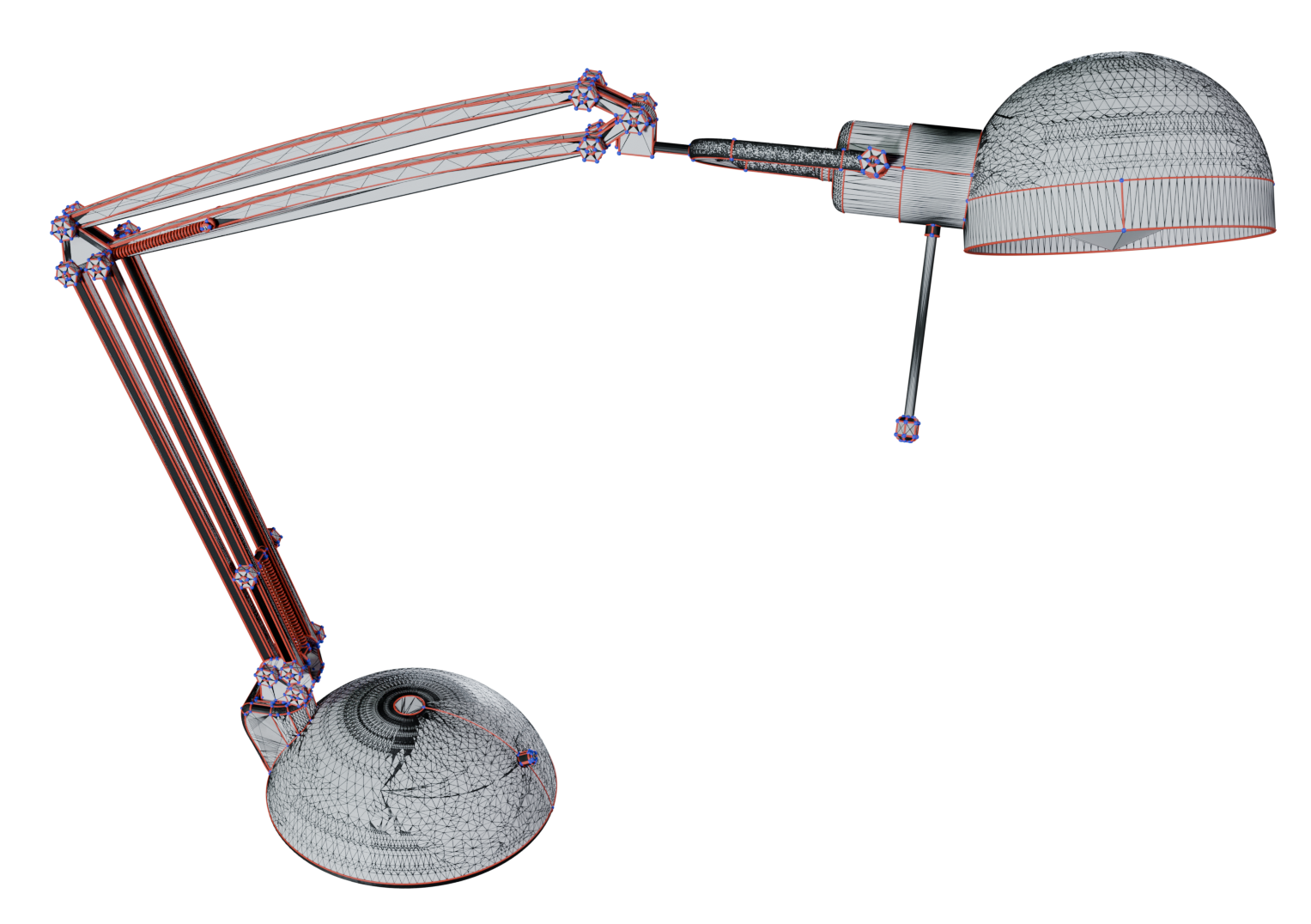}}\hfill
\parbox{.18\linewidth}{\includegraphics[width=\linewidth]{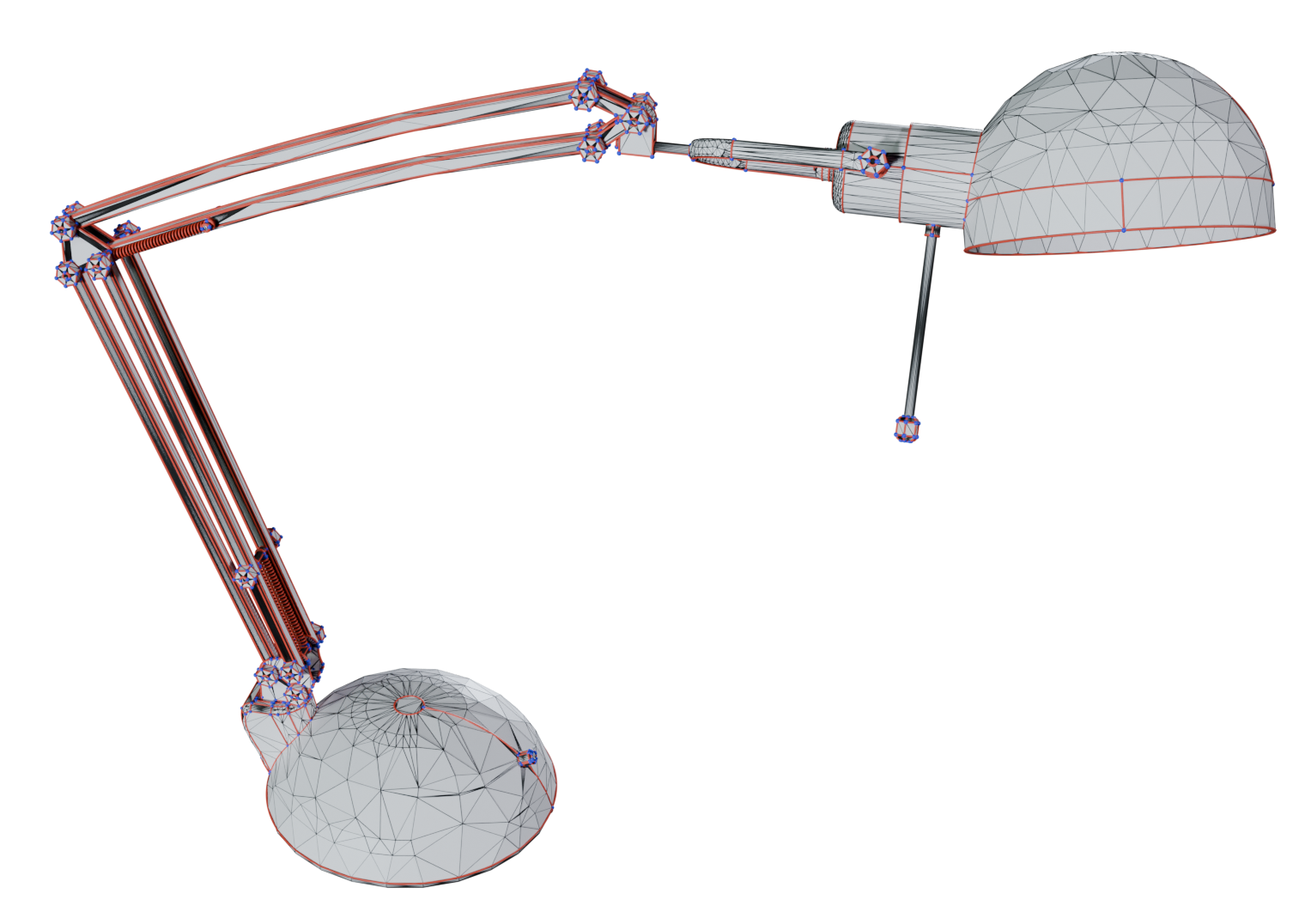}}\hfill
\parbox{.18\linewidth}{\centering Crash}\hfill
\parbox{.18\linewidth}{\includegraphics[width=\linewidth]{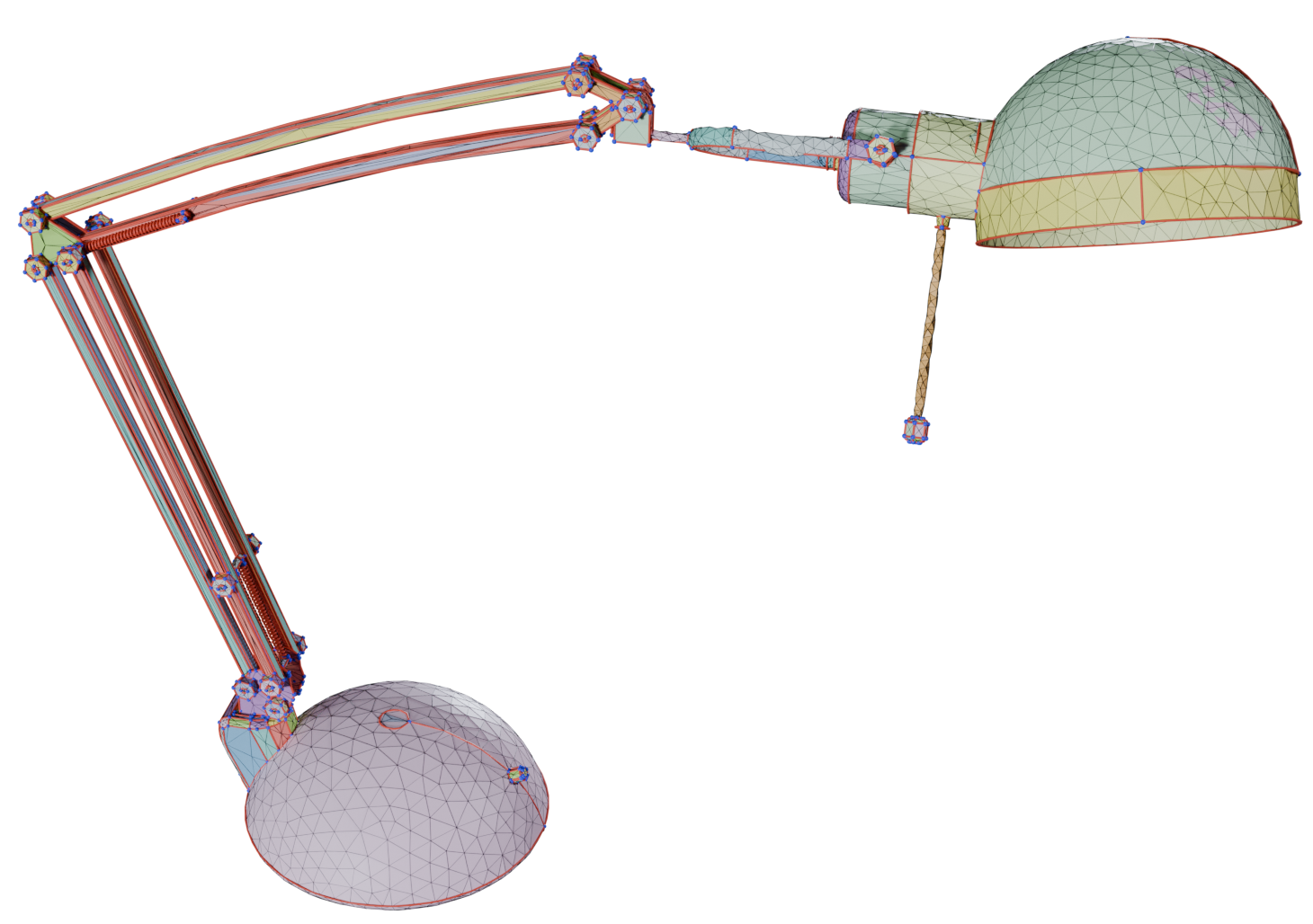}}

\parbox{.01\linewidth}{\rotatebox{90}{\centering Default}}\hfill
\parbox{.18\linewidth}{\centering Out of Memory}\hfill
\parbox{.18\linewidth}{\includegraphics[width=\linewidth]{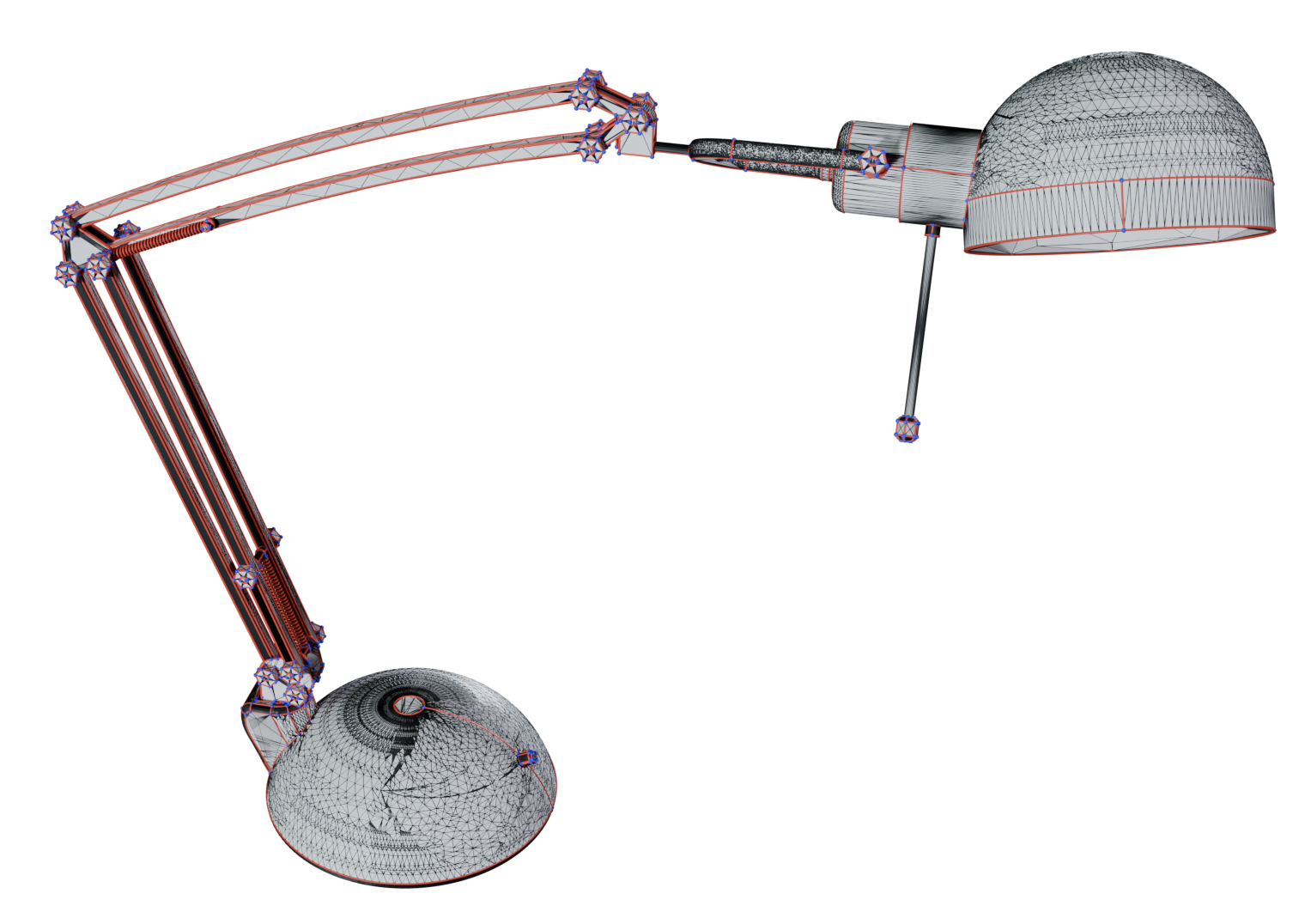}}\hfill
\parbox{.18\linewidth}{\includegraphics[width=\linewidth]{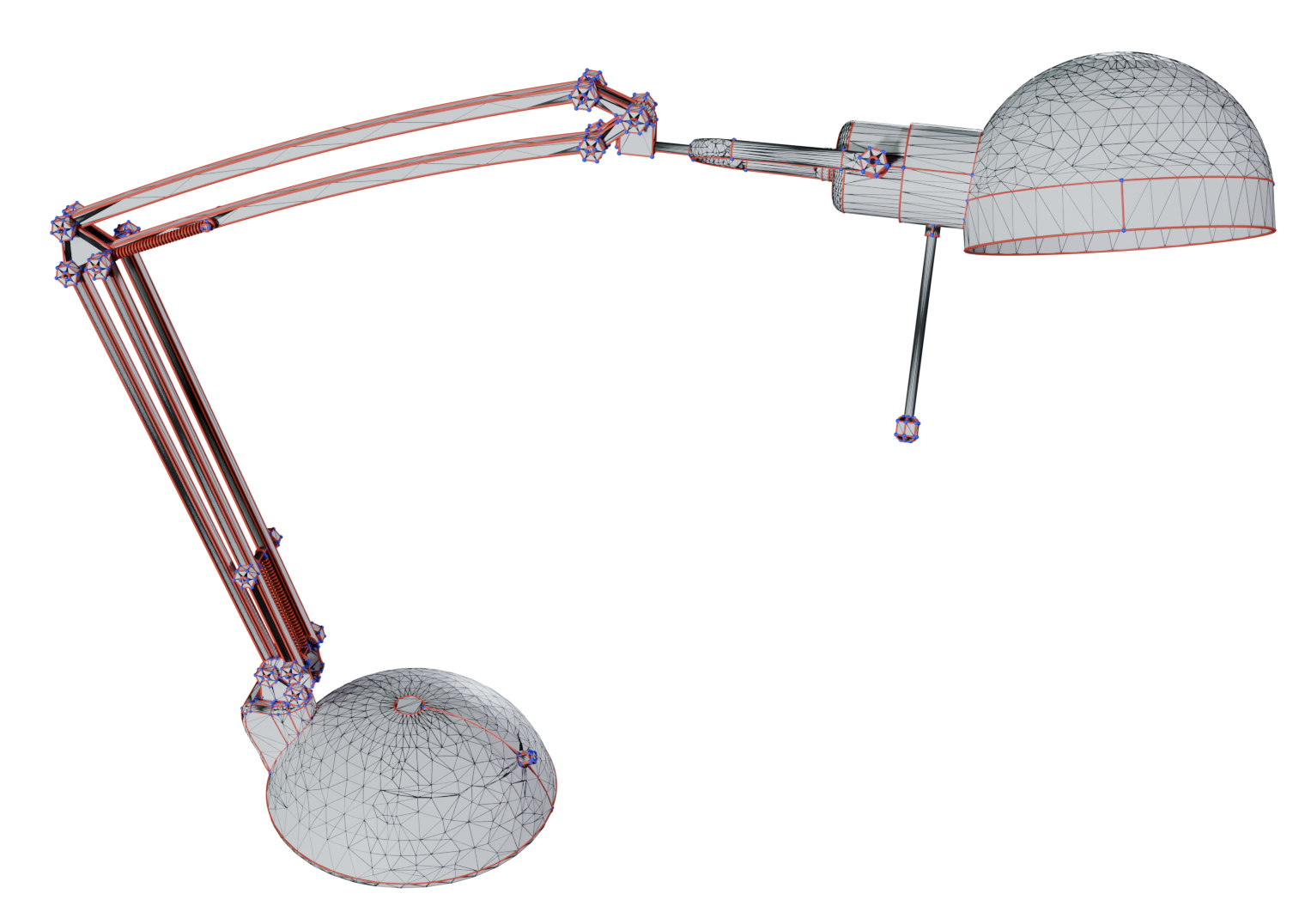}}\hfill
\parbox{.18\linewidth}{\centering Crash}\hfill
\parbox{.18\linewidth}{\includegraphics[width=\linewidth]{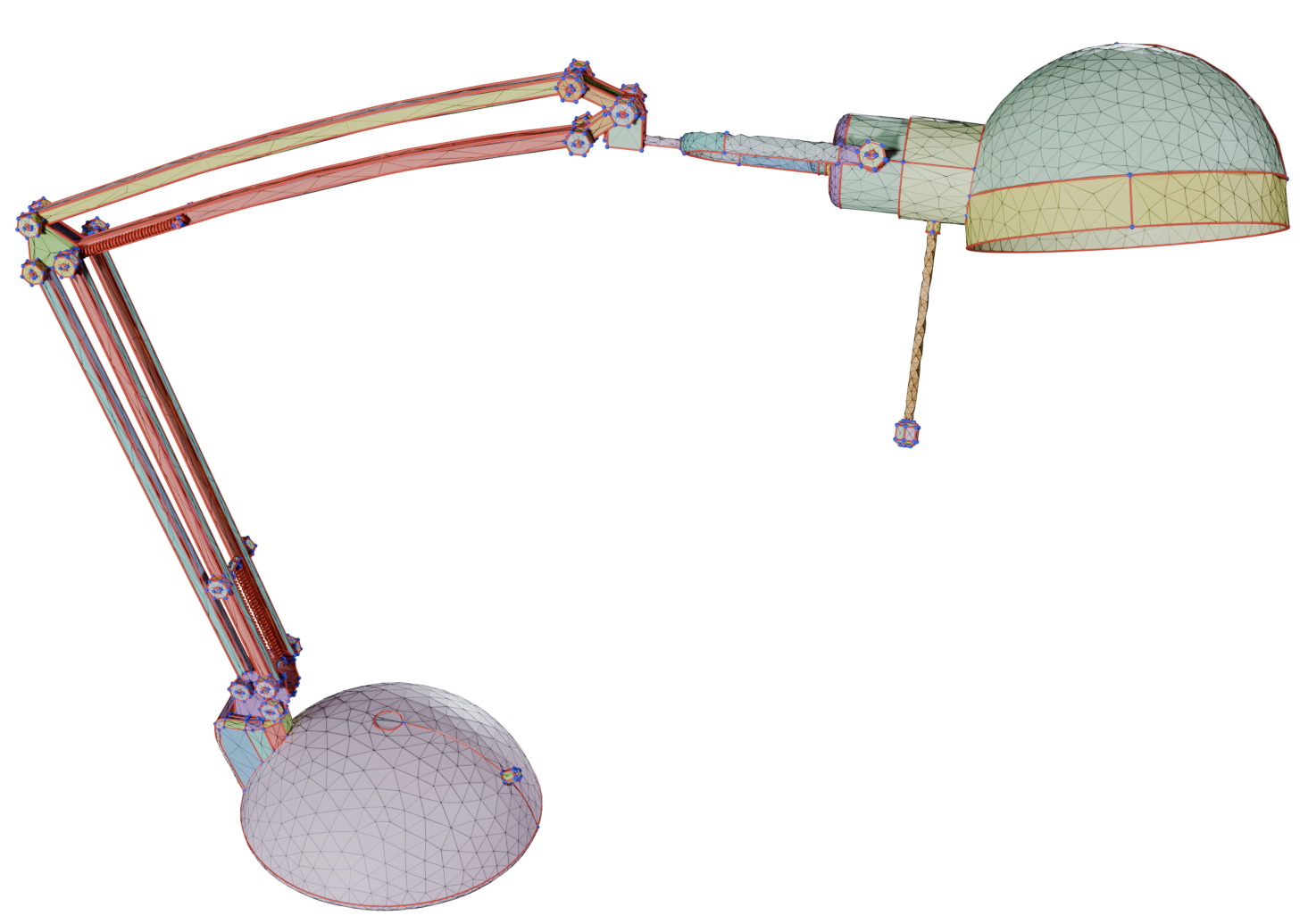}}

\parbox{.01\linewidth}{\rotatebox{90}{\centering Fine}}\hfill
\parbox{.18\linewidth}{\centering Out of Memory}\hfill
\parbox{.18\linewidth}{\includegraphics[width=\linewidth]{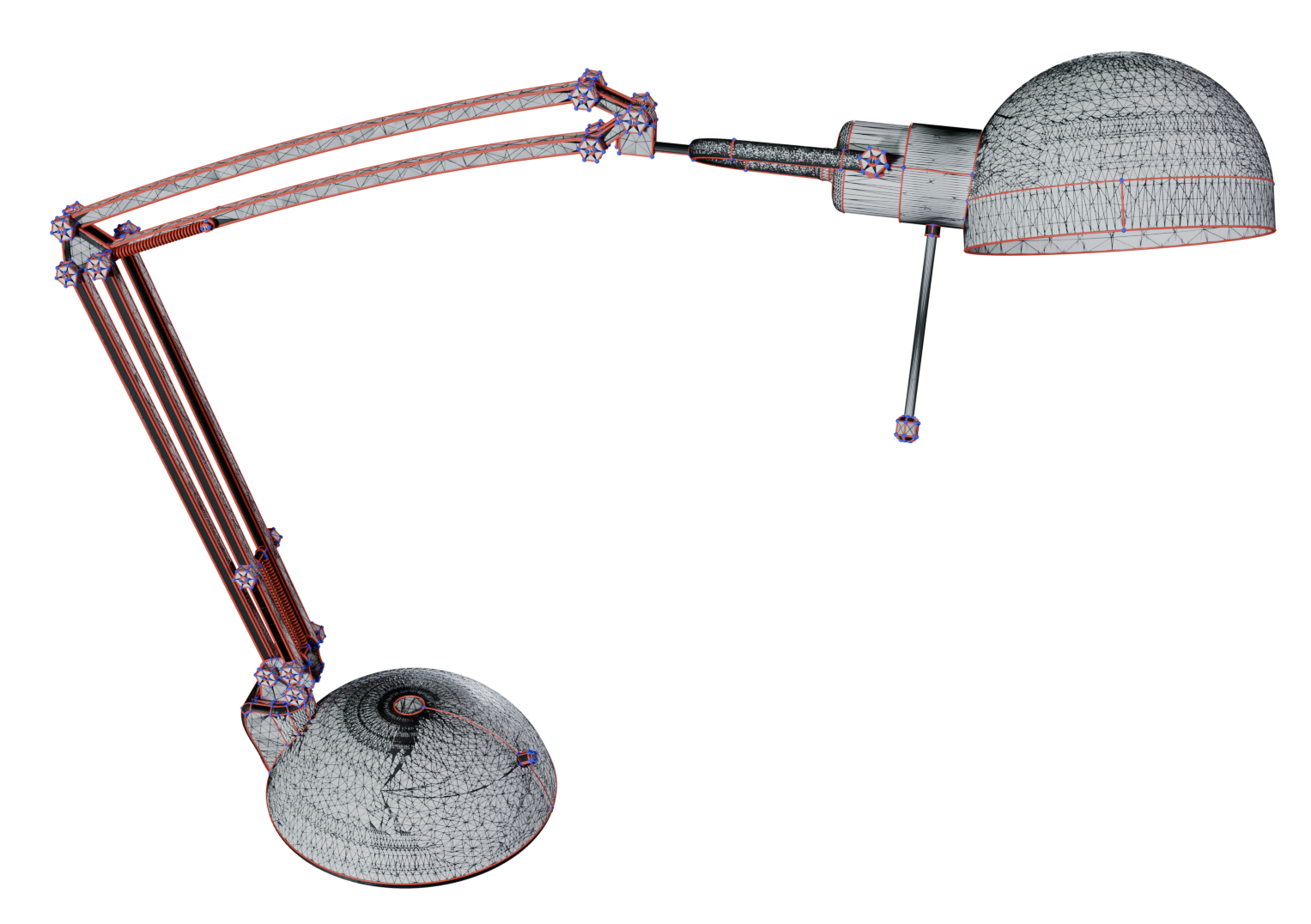}}\hfill
\parbox{.18\linewidth}{\includegraphics[width=\linewidth]{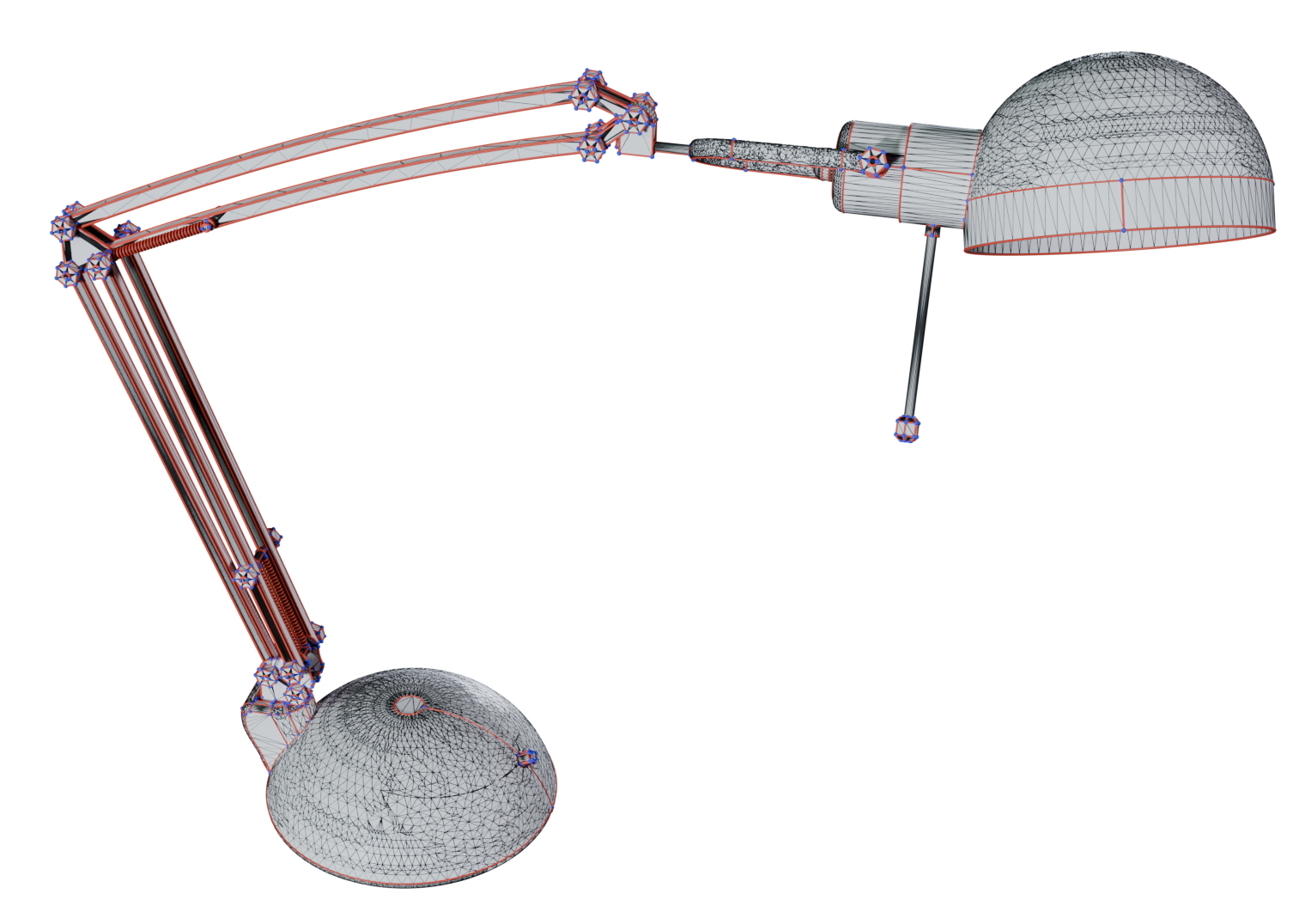}}\hfill
\parbox{.18\linewidth}{\centering Crash}\hfill
\parbox{.18\linewidth}{\includegraphics[width=\linewidth]{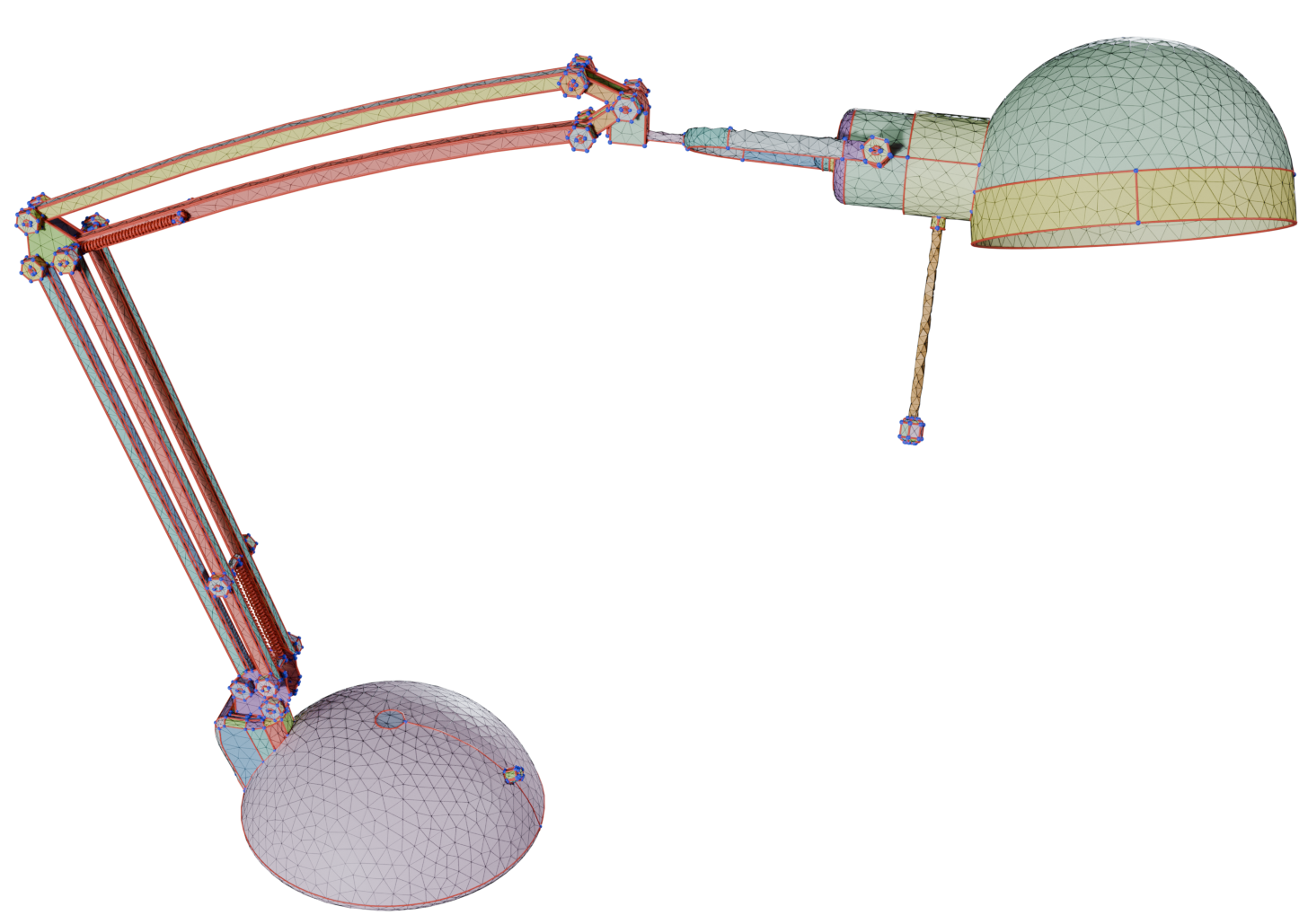}}

\caption{
Comparison of B-Rep meshing results across different methods and resolutions.
Columns correspond to Gmsh, Mefisto, OpenCascade (OCC), NetGen, and our method.
Rows show three parameter regimes (coarse, default, and fine).
Failed runs are explicitly indicated.
}
\label{fig:comparison}
\end{figure*}

\begin{figure*}[t]\ContinuedFloat

\parbox{.01\linewidth}{~}\hfill
\parbox{.18\linewidth}{\centering gmsh}\hfill
\parbox{.18\linewidth}{\centering Mefisto}\hfill
\parbox{.18\linewidth}{\centering OCC}\hfill
\parbox{.18\linewidth}{\centering NetGen}\hfill
\parbox{.18\linewidth}{\centering Ours}






\parbox{.01\linewidth}{\rotatebox{90}{\centering Coarse}}\hfill
\parbox{.18\linewidth}{\includegraphics[width=\linewidth]{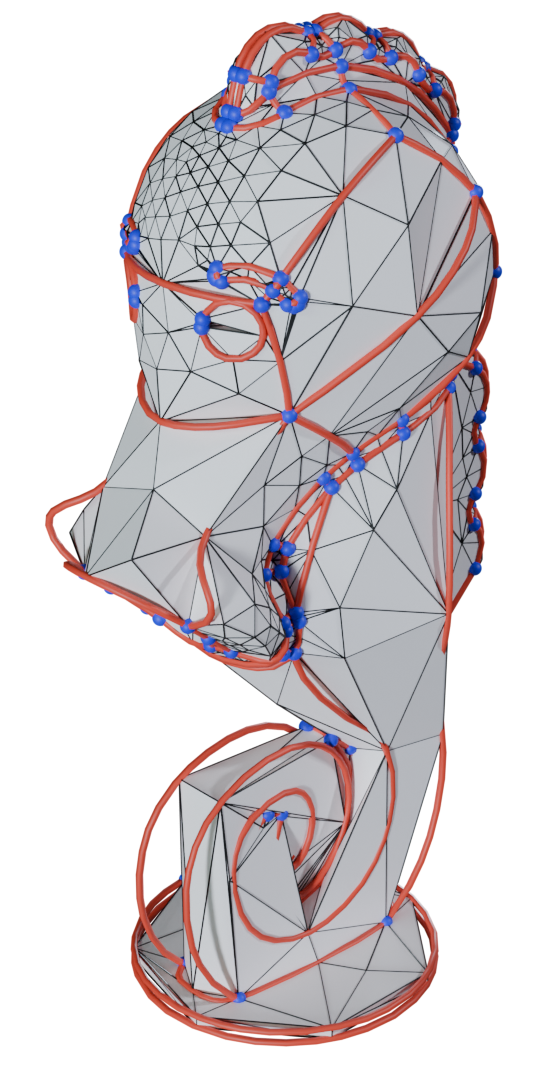}}\hfill
\parbox{.18\linewidth}{\includegraphics[width=\linewidth]{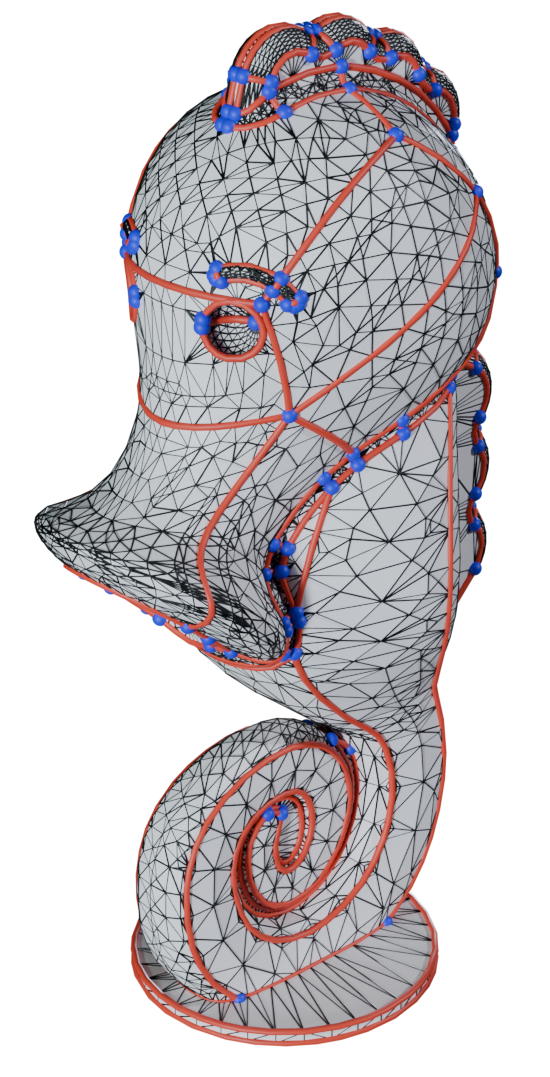}}\hfill
\parbox{.18\linewidth}{\includegraphics[width=\linewidth]{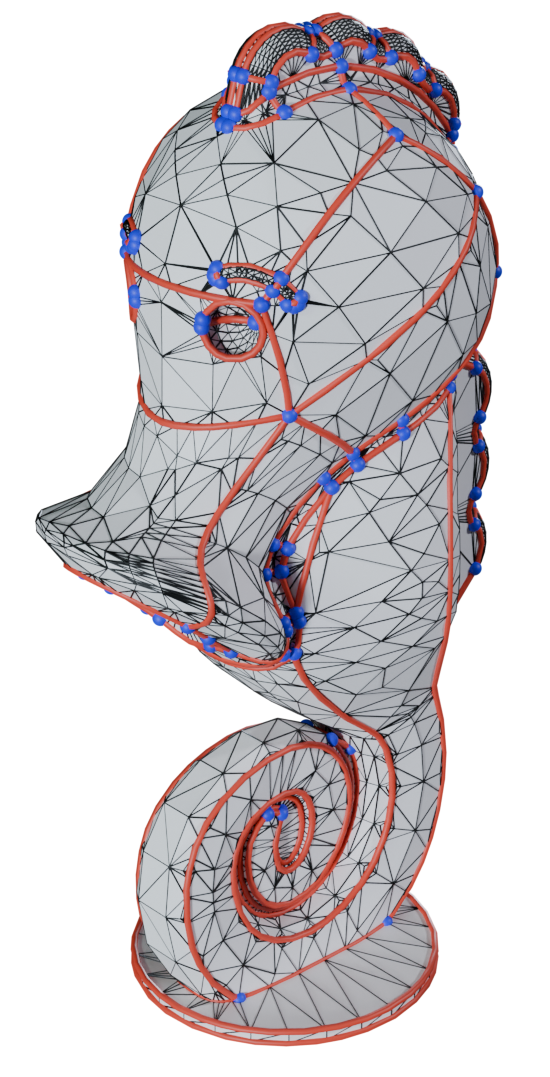}}\hfill
\parbox{.18\linewidth}{\includegraphics[width=\linewidth]{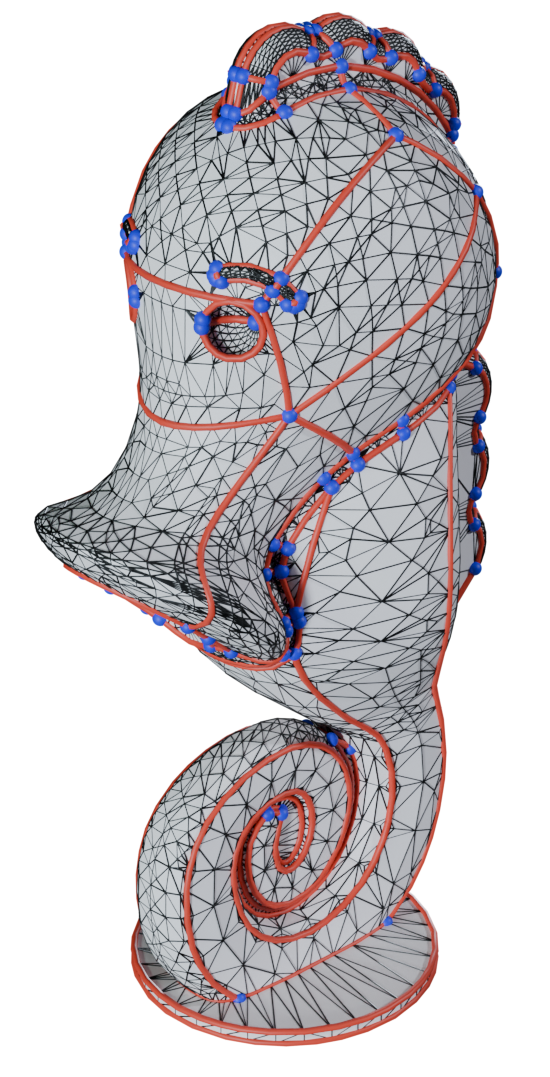}}\hfill
\parbox{.18\linewidth}{\includegraphics[width=\linewidth]{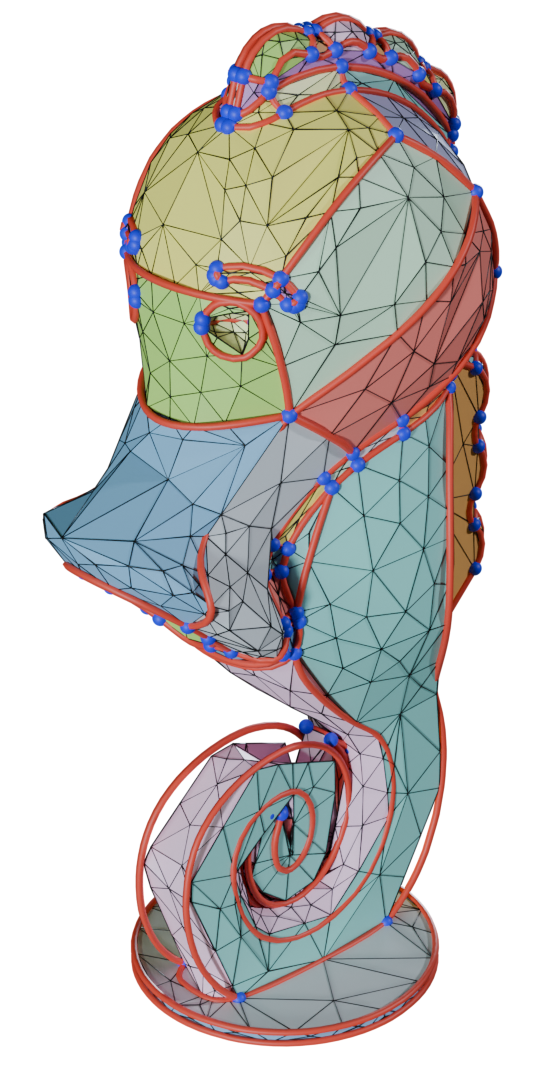}}

\parbox{.01\linewidth}{\rotatebox{90}{\centering Default}}\hfill
\parbox{.18\linewidth}{\includegraphics[width=\linewidth]{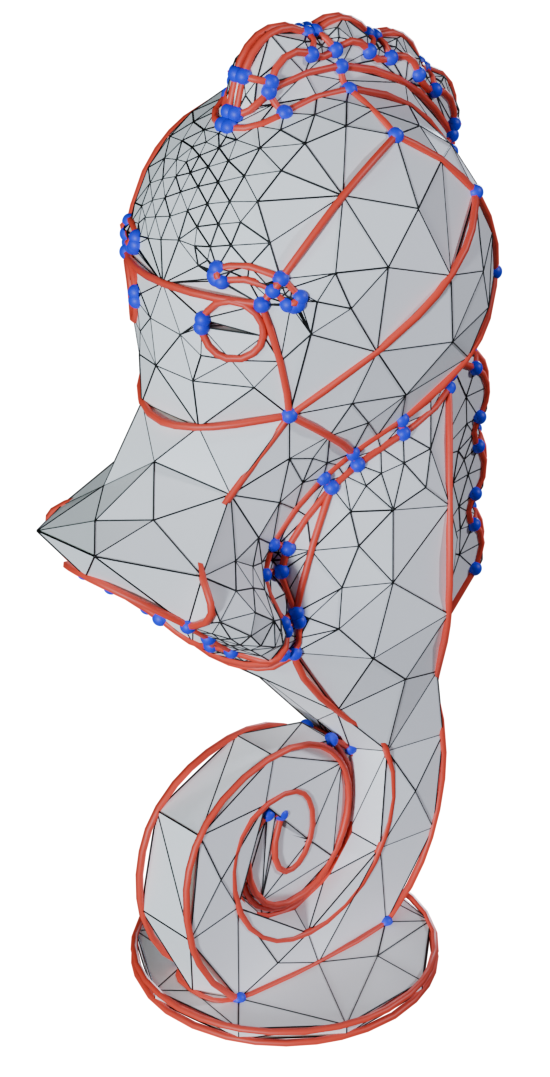}}\hfill
\parbox{.18\linewidth}{\includegraphics[width=\linewidth]{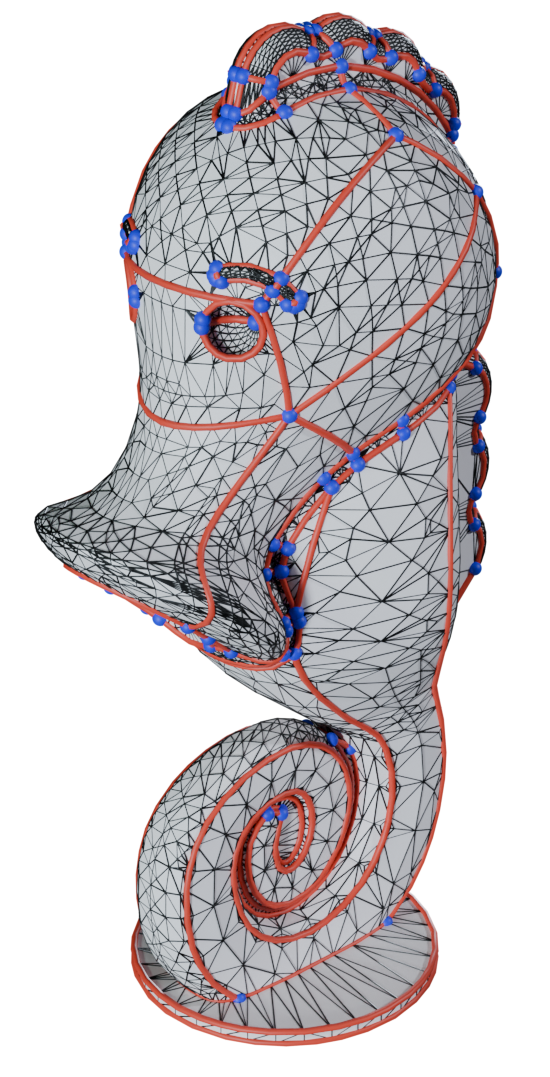}}\hfill
\parbox{.18\linewidth}{\includegraphics[width=\linewidth]{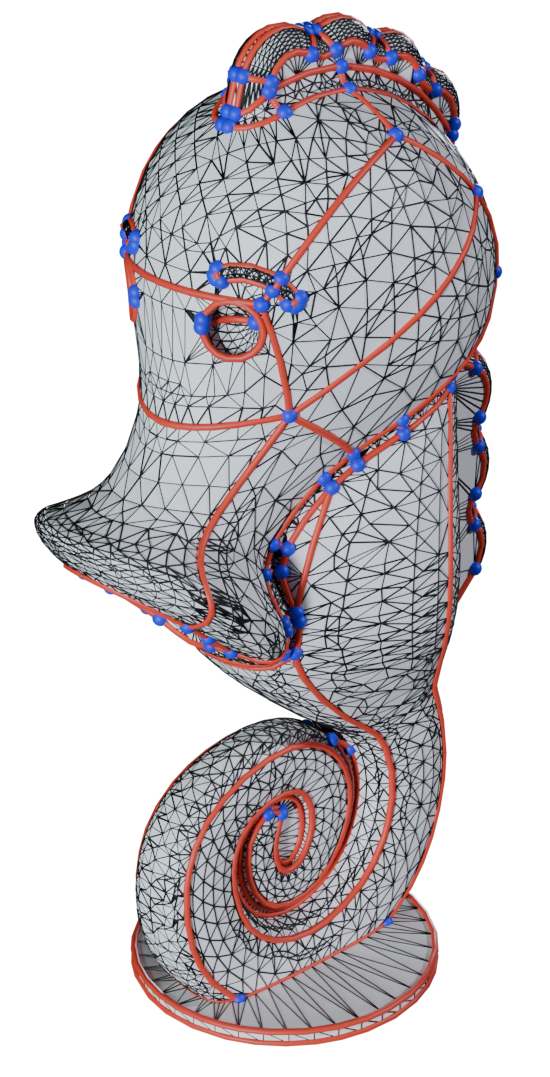}}\hfill
\parbox{.18\linewidth}{\includegraphics[width=\linewidth]{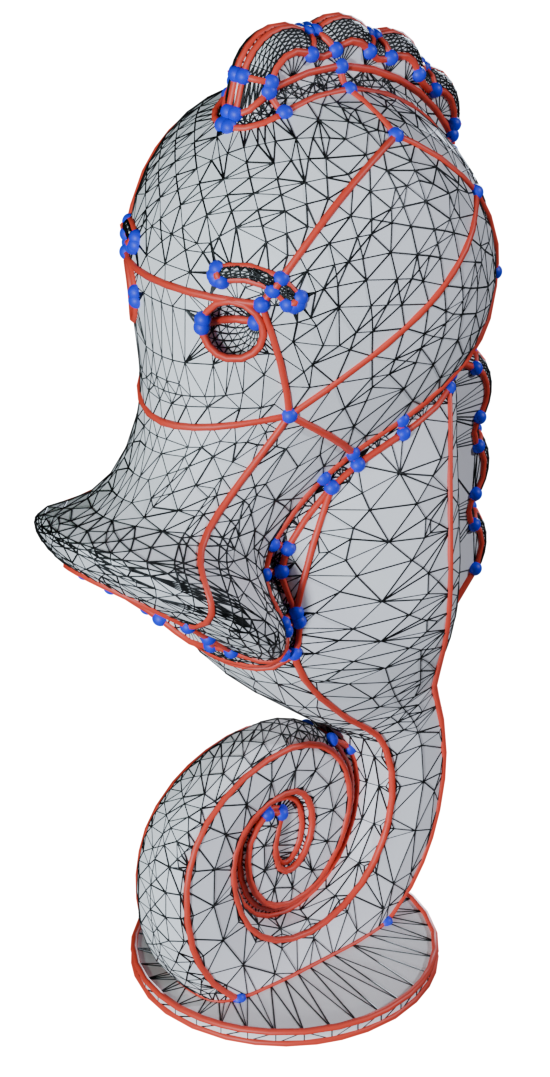}}\hfill
\parbox{.18\linewidth}{\includegraphics[width=\linewidth]{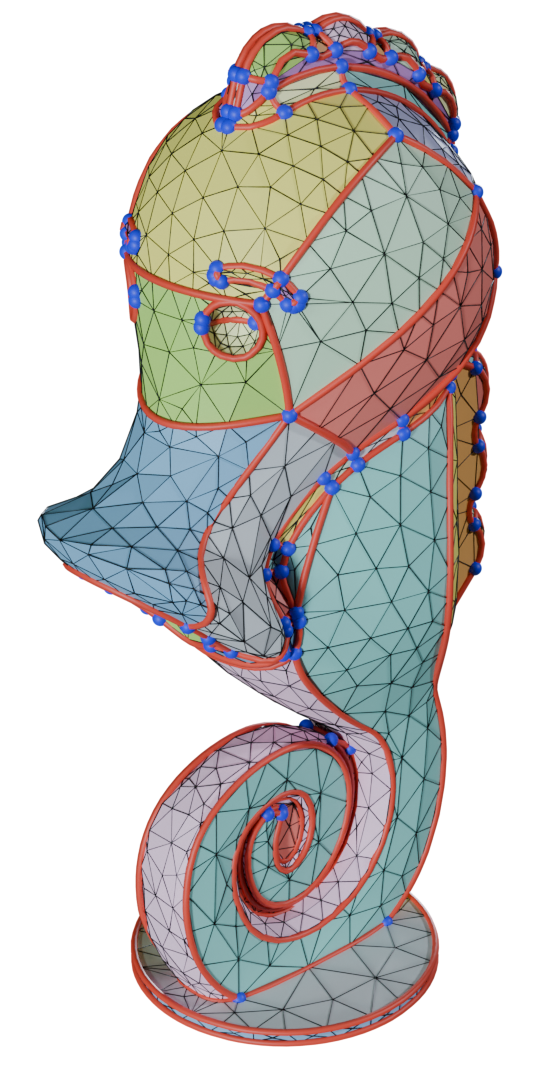}}

\parbox{.01\linewidth}{\rotatebox{90}{\centering Fine}}\hfill
\parbox{.18\linewidth}{\includegraphics[width=\linewidth]{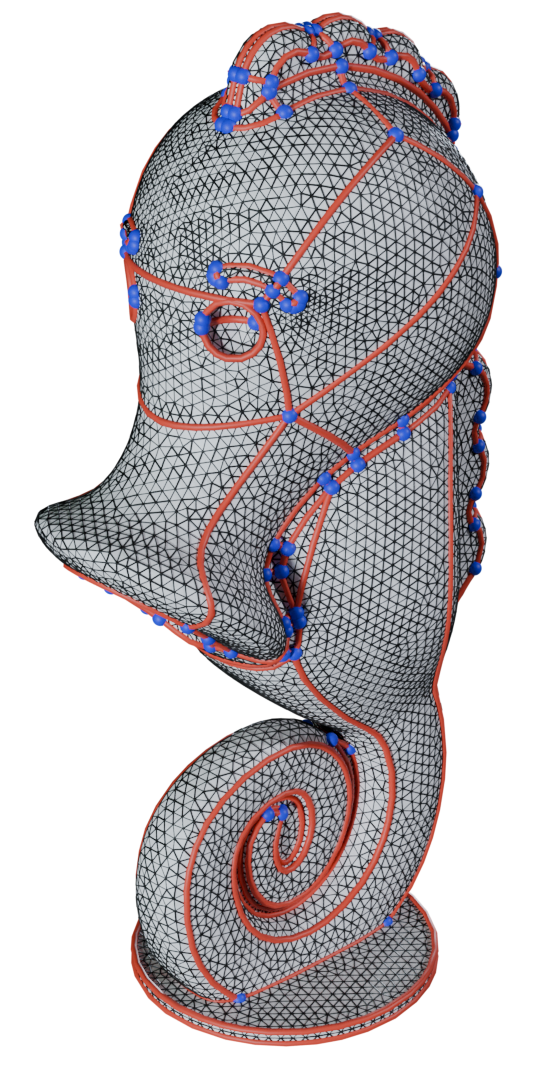}}\hfill
\parbox{.18\linewidth}{\includegraphics[width=\linewidth]{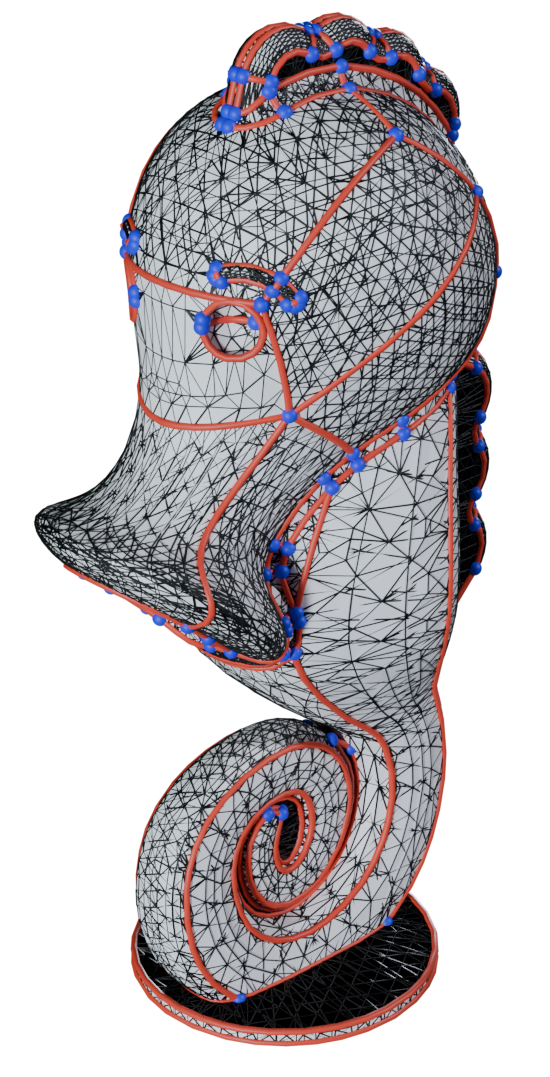}}\hfill
\parbox{.18\linewidth}{\includegraphics[width=\linewidth]{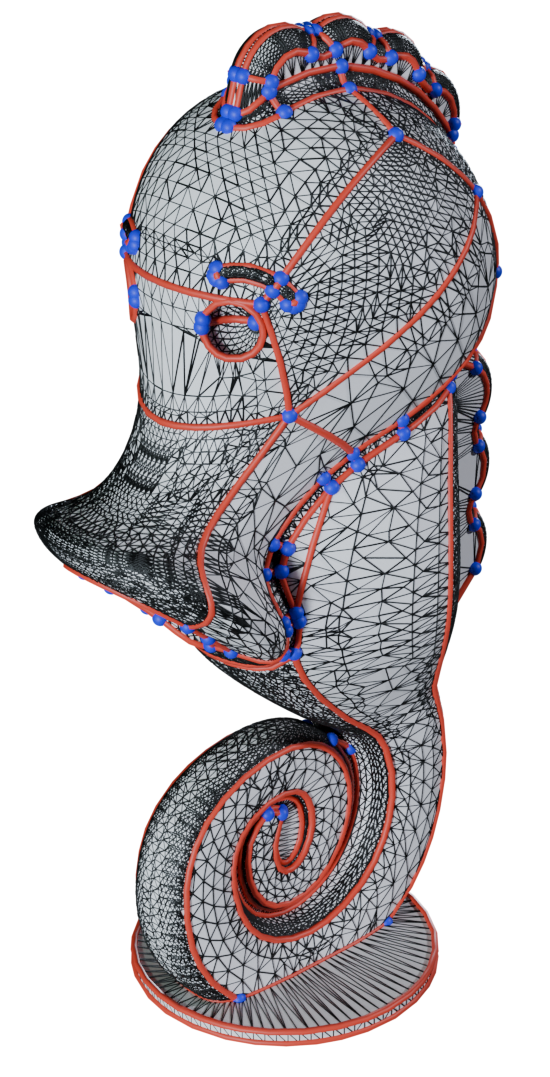}}\hfill
\parbox{.18\linewidth}{\includegraphics[width=\linewidth]{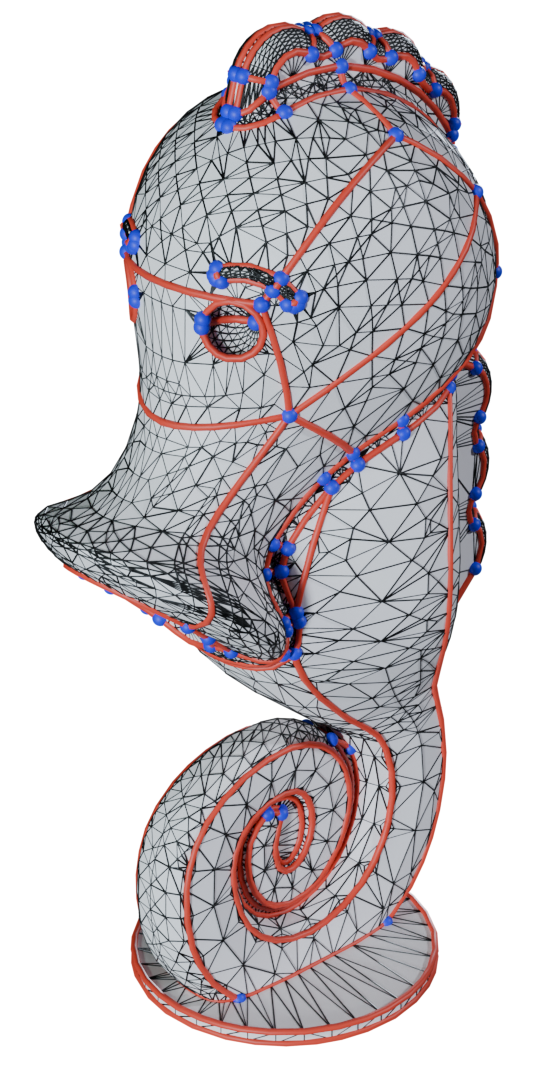}}\hfill
\parbox{.18\linewidth}{\includegraphics[width=\linewidth]{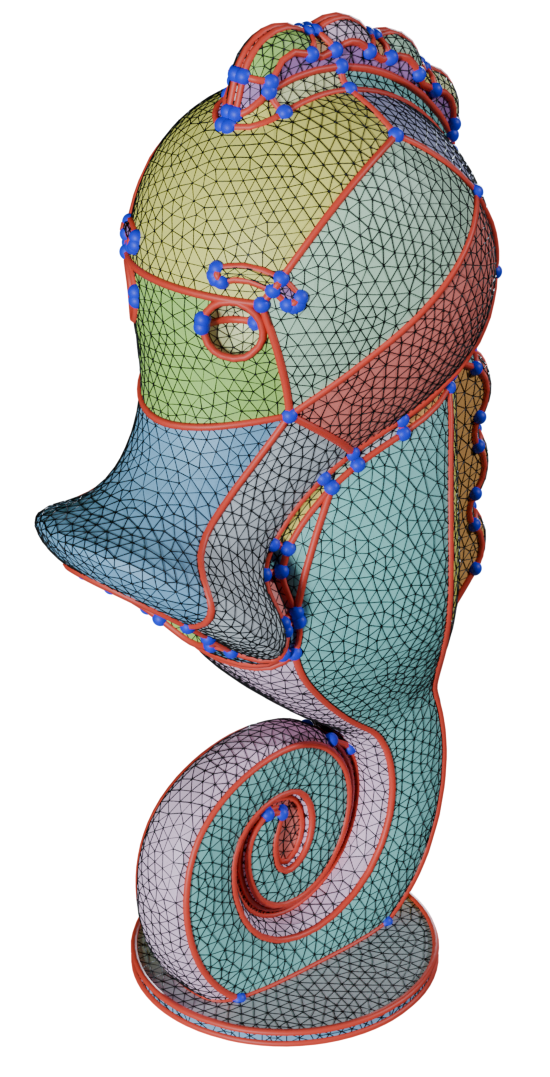}}


\end{figure*}

%% file: 60-conclusion.tex
\section{Conclusion}
\label{sec:conclusion}

We presented a novel approach to convert B-Reps into triangle meshes that preserves the B-Rep topology. We compared with existing algorithms, demonstrating their superior reliability and consistent quality in the generated meshes.

These properties lead to higher computational cost than competing methods: we believe this is a worthwhile tradeoff for use cases that favor automation over running time.

There are two main avenues of work that we believe are interesting: (1) we wonder if similar topology first approaches could also be applied to the CAD kernel itself, where the current numerical methods used to compute intersections and trims could be complemented by a stronger topological prior (for example, the genus of the intersection between two patches has to be 1) to obtain more reliably implementation of CAD kernel operations. (2) We currently only track correspondences between the edge features, but it would be possible to propagate this information to the patch interiors using a bijective parametrization approach, to obtain dense correspondences between the B-Rep and its mesh counterpart. This could be useful for inverse design algorithms or for transferring physical quantities computed on the triangle mesh back to the B-Rep.

Although our method provides strong topological guarantees, it offers only best-effort preservation of geometry and does not provide formal geometric guarantees, even if the observed errors are small in practice (\Cref{fig:geom-error}).
If the input geometry is inconsistent or ill-posed, our algorithm will still produce a mesh, but the resulting geometry may be poor.
Investigating how to integrate explicit geometric guarantees into our pipeline is an important direction for future work.
Similarly, when the input topology is itself pathological, we do not attempt to repair it, but instead, we faithfully preserve it.

%% file: 90-appendix.tex
\crefalias{section}{appsec}